\documentclass{elsart}

\usepackage{graphics}

\usepackage{amssymb}

\usepackage{array}

\newcommand{\jpsi}{J/\psi}
\newcommand{\ks}{K_S^0}
\newcommand{\kl}{K_L^0}
\newcommand{\kstarz}{K^{*0}}
\newcommand{\etac}{\eta_{c}}
\newcommand{\bz}{B^{0}}
\newcommand{\bzb}{\overline{B}{}^0}
\newcommand{\dedx}{${\rm d}E/{\rm d}x$}

\newcommand{\dmd}{\Delta m_d}
\newcommand{\Dt}{\Delta t}
\newcommand{\fcp}{f_{CP}}

\newcommand{\taubz}{\tau_{\bz}}

\begin{document}

\begin{frontmatter}



\title{Neutral $B$ Flavor Tagging for the Measurement of Mixing-induced
  $CP$ Violation at Belle}


\author[TIT]{H. Kakuno\corauthref{cor1}},
\ead{kakuno@hp.phys.titech.ac.jp}
\corauth[cor1]{Corresponding author.
  Tel: +81-3-5734-2388; fax: +81-3-5734-2389}
\author[Osaka]{K. Hara},
\author[Hawaii]{B. C. K. Casey},
\author[NTU]{K.-F. Chen},
\author[KEK]{H. Hamasaki},
\author[KEK]{M. Hazumi},
\author[Nagoya]{T. Iijima},
\author[KEK]{N. Katayama},
\author[Nagoya]{T. Okabe},
\author[KEK]{Y. Sakai},
\author[KEK]{K. Sumisawa},
\author[KEK]{J. Suzuki},
\author[Tokyo]{T. Tomura},
\author[NTU]{K. Ueno},
\author[NTU]{C.-C. Wang},
\author[TIT]{and Y. Watanabe}

\address[TIT]{Department of Physics, Tokyo Institute of Technology,
  Tokyo 152-8551}
\address[Osaka]{Department of Physics, Osaka University, Osaka
  560-0043, Japan}
\address[Hawaii]{University of Hawaii, Hawaii, USA}
\address[NTU]{National Taiwan University, Taipei}
\address[KEK]{High Energy Accelerator Organization (KEK), Tsukuba
  305-0801, Japan}
\address[Nagoya]{Nagoya University, Nagoya, Japan}
\address[Tokyo]{Department of Physics, University of Tokyo, Tokyo
  113-0033, Japan}

\begin{abstract}
We describe a flavor tagging algorithm used in
measurements of the $CP$ violation parameter $\sin2 \phi_1$
at the Belle experiment.
Efficiencies and wrong tag fractions are evaluated using
flavor-specific $B$ meson decays into hadronic and semileptonic modes.
We achieve a total effective efficiency of $ 28.8 \pm 0.6 \%$.
\end{abstract}


\end{frontmatter}

\section{Introduction}
\label{sec:intro}
In the Standard Model (SM) of elementary particles, $CP$ violation
arises from an irreducible complex phase in the weak interaction
quark-mixing matrix (CKM matrix)~\cite{bib:ckm}.
In particular, the SM predicts a $CP$-violating asymmetry
in the time-dependent rates for $\bz$ and $\bzb$ decays
to a common $CP$ eigenstate $\fcp$~\cite{bib:sanda}.
This $CP$-violating asymmetry in the $\fcp$ dominated by the
$b \to c\overline{c}s$ transition, has recently been observed
by the Belle and BaBar groups~\cite{bib:cpv,bib:babar}.
The measurement at Belle is based on a sample of $B\overline{B}$ pairs
collected at the $\Upsilon(4S)$ resonance at the KEKB
asymmetric-energy $e^+e^-$ collider.
In the decay chain $\Upsilon(4S)\to \bz\bzb \to f_{CP}f_{\rm tag}$,
where one of the two $B$ mesons decays at time $t_{CP}$ to $f_{CP}$
and the other, at time $t_{\rm tag}$ to a final state $f_{\rm tag}$
that distinguishes $B^0$ and $\bzb$,
the decay rates and their asymmetry are time dependent.
The time dependence for $b \rightarrow c\bar{c}s$ transitions is given by
\begin{eqnarray}
  \label{eq:deltat}
  {\mathcal P}_{\rm sig} (\Dt, q, \xi_f) & = &
  \frac{e^{-|\Dt|/\taubz}}{4\taubz}[1 - q\xi_f\sin
  2\phi_1\sin(\dmd\Dt)], \\
{\mathcal A}_{CP} & = & \frac{{\mathcal P}_{\rm sig}(\Dt, q,
\xi_f) - {\mathcal P}_{\rm sig}(\Dt, -q, \xi_f)}
{{\mathcal P}_{\rm sig}(\Dt, q, \xi_f) +
{\mathcal P}_{\rm sig}(\Dt, -q, \xi_f)}.
\end{eqnarray}
where ${\mathcal P}_{\rm sig}$ represents the normalized decay rate,
$\taubz$ is the $B^0$ lifetime, $\xi_f$ is the CP-eigenvalue of
$f_{CP}$, $\dmd$ is the mass difference between the two $B^0$ mass
eigenstates, $\Delta{t}$ = $t_{CP}$ $-$ $t_{\rm tag}$, the $b$-flavor
charge $q$ = +1 ($-1$) when the tagging $B$ meson is a $B^0$ ($\bzb$).
The $CP$ parameter $\phi_1$ is one of the three interior angles of the CKM
unitarity triangle,
defined as $\phi_1 \equiv \pi - \arg(V_{tb}^*V_{td}/V_{cb}^*V_{cd})$.
The $CP$ eigenstates are reconstructed from
$B \rightarrow \jpsi\ks$, $\psi(2S)\ks$, $\chi_{c1}\ks$, $\etac\ks$,
$\jpsi\kstarz(\kstarz \rightarrow \ks \pi^0)$, or $\jpsi\kl$ decays.

An identification of the flavor of the accompanying $B$ meson,
called flavor tagging in this article, is required to observe this
kind of $CP$-violating asymmetry.
A perfect tagging algorithm with a perfect detector will tag
every $B$ meson that decays into a flavor specific decay mode and
will identify the flavor of the $B$ meson.
A practical tagging algorithm in a realistic detector will tag only
a fraction $\epsilon$ (the tagging efficiency) of $B$ mesons and
of those tagged, only a fraction of them will be identified
correctly. The fraction of $B$ mesons identified incorrectly is
called the wrong tag fraction $w$.
The observed time dependence ${\mathcal P}_{\rm sig}^{\rm obs}$
thus becomes
\begin{eqnarray}
\label{eq:deltat2}
  {\mathcal P}_{\rm sig}^{\rm obs} (\Dt, q, w, \xi_f) & = &
  \epsilon\cdot((1-w) {\mathcal P}_{\rm sig} (\Dt, q, \xi_f)
  + w {\mathcal P}_{\rm sig} (\Dt, -q, \xi_f)) \nonumber \\
  & = & \epsilon\cdot
  \frac{e^{-|\Dt|/\taubz}}{4\taubz}[1 - (1-2w)q\xi_f\sin 2\phi_1\sin(\dmd\Dt)],
\end{eqnarray}
and the observed $CP$-violating asymmetry ${\mathcal A}_{CP}^{\rm obs}$,
\begin{eqnarray}
  {\mathcal A}_{CP}^{\rm obs}
  & = & \frac{{\mathcal P}_{\rm sig}^{\rm obs}(\Dt, q, w,
  \xi_f) - {\mathcal P}_{\rm sig}^{\rm obs}(\Dt, -q, w, \xi_f)}
  {{\mathcal P}_{\rm sig}^{\rm obs}(\Dt, q, w,\xi_f) +
  {\mathcal P}_{\rm sig}^{\rm obs}(\Dt, -q, w, \xi_f)} \nonumber \\
  & = & -(1-2w)q\xi_f\sin 2\phi_1\sin(\dmd\Dt) = (1-2w){\mathcal A}_{CP}.
\end{eqnarray}
Here, we ignore a possible small difference between wrong tag fraction
for $q=+1$ and $q=-1$.
The observed $CP$-violating asymmetry ${\mathcal A}_{CP}^{\rm obs}$
is diluted by $(1-2w)$, which is called the dilution factor.
The statistical significance of the asymmetry measurement is proportional
to $(1-2w)\sqrt{\epsilon}$, {\it i.e.} the number of events required to
observe the asymmetry for a certain statistical significance is
inversely proportional to $\epsilon_{\rm eff} \equiv \epsilon (1-2w)^2$,
which is called the ``effective tagging efficiency''.
At the same time, since the the factor $(1-2w)\sin 2\phi_1$ is
proportional to the amplitude of observed $CP$-asymmetry,
the wrong tag fraction $w$ directly affects the central value
of $\sin 2\phi_1$.
Therefore, a precise measurement of $w$ is crucial in order to minimize
the systematic uncertainty in the $\sin 2\phi_1$ measurement.

Our tagging algorithm has been developed to maximize $\epsilon_{\rm eff}$
while making it possible to determine the value of $w$ experimentally.
The flavor tagging described in this paper is used not only for the
$\sin 2\phi_1$ measurement but also for other measurements such as
$\Delta m_d$ measurements~\cite{dslnu_mixing,hadron_mixing} and
measurements of $CP$-violating asymmetries in the
decay $B\rightarrow \pi^+ \pi^-$~\cite{a_pipi}.
In this paper, we present an algorithm for flavor tagging and describe its
performance. The experimental apparatus of the Belle experiment is
described in the next section. The flavor tagging algorithm is
described in Section~\ref{sec:algorithm}, and the measurement of its
performance and wrong tag fractions with the control samples
of self-tagged neutral $B$ decays, in Section~\ref{sec:performance}.

\section{Experimental Apparatus}
\label{sec:belle}
The Belle experiment is conducted at the KEKB energy-asymmetric
$e^{+}$ (3.5~GeV) $e^{-}$ (8.0~GeV) collider with a crossing angle of 22 mrad.
The corresponding center-of-mass (CMS) energy is 10.58~${\rm GeV}$,
which is on the $\Upsilon(4S)$ resonance.
The $\Upsilon(4S)$ decays into $B\bar{B}$ pairs with a Lorentz boost of
$(\beta \gamma)_{\Upsilon(4S)} = 0.425$ nearly along the $z$ axis,
which is defined as opposite to the positron beam direction.
The time difference between the two $B$ meson decays is measured from the
distance between the two $B$ decay vertices ($\Dt = \Delta z/\beta \gamma c$).

The Belle detector \cite{cite:belle_detector} is a general-purpose
spectrometer surrounding the interaction point.
It consists of a barrel, forward and backward components.
It is placed in such a way that the axis of the detector solenoid is
parallel to the $z$ axis.
In this way, the Lorentz force on the low energy positron beam is minimized.

Precision tracking and vertex measurements are provided by a central
drift chamber (CDC)~\cite{CDC} and a silicon vertex detector (SVD)~\cite{SVD}.
The CDC is a small-cell cylindrical drift chamber with 50 layers of
anode wires including 18 layers of stereo wires. 
A low-$Z$ gas mixture [He (50\%) and ${\rm C}_2{\rm H}_6$ (50\%)] is used
to minimize multiple Coulomb scattering and to ensure a good momentum
resolution, especially for low momentum particles.
It provides three-dimensional trajectories of charged particles
in the polar angle region $17^\circ < \theta < 150^\circ$ in the
laboratory frame, where $\theta$ is measured with respect to the $z$ axis.
The SVD consists of three layers of double-sided silicon strip detectors
arranged in a barrel and covers 86\% of the solid angle. The three layers
at radii of 3.0, 4.5 and 6.0 cm surround the beam-pipe, a double-wall
beryllium cylinder of 2.3~cm outer radius and 1~mm  thickness.
The strip pitches are 84~$\mu$m for the measurement of $z$
coordinate and 25~$\mu$m for the measurement of the azimuthal angle $\phi$.
The impact parameter resolution for reconstructed tracks is measured as
a function of the track momentum $p$ (measured in GeV/{\it c}) to be
$\sigma_{xy}$ = [19 $\oplus$ 50/($p\beta\sin^{3/2}\theta$)]~$\mu$m and
$\sigma_{z}$ = [36 $\oplus$ 42/($p\beta\sin^{5/2}\theta $)]~$\mu$m.
The momentum resolution of the combined tracking system is
$\sigma_{p_{\rm t}}/p_{\rm t} = (0.30/\beta \oplus 0.19p_{\rm t})$\%,
where $p_{\rm t}$ is the transverse momentum in GeV/{\it c}.

The identification of charged pions and kaons uses three detector systems:
the CDC measurements of \dedx, a set of time-of-flight counters (TOF)\cite{TOF}
and a set of aerogel \v{C}herenkov counters (ACC)\cite{ACC}.
The CDC measures energy loss for charged particles with a resolution of
$\sigma({\rm d}E/{\rm d}x)$ = 6.9\% for minimum-ionizing pions.
The TOF consists of 128 plastic scintillators viewed on both ends by fine-mesh
photo-multipliers that operate stably in the 1.5~T magnetic field.
Their time resolution is 95~ps ($rms$) for minimum-ionizing particles,
providing three standard deviation (3$\sigma$) $K^\pm/\pi^\pm$ separation
below 1.0~GeV/$c$, and 2$\sigma$ up to 1.5~GeV/$c$.
The ACC consists of 1188 aerogel blocks with refractive indices between
1.01 and 1.03 depending on the polar angle.
Fine-mesh photo-multipliers detect the \v{C}herenkov light.
The effective number of photoelectrons is approximately 6 for $\beta =1$
particles.
Using this information, $P(K/\pi) = Prob(K)/(Prob(K)+Prob(\pi))$,
the probability for a particle to be a $K^\pm$ meson, is calculated.
A selection with $P(K/\pi) > 0.6$ retains about 90\% of the charged kaons
with a charged pion misidentification rate of about 6\%.

Photons are reconstructed in a CsI(Tl) crystal calorimeter (ECL)~\cite{ECL}
consisting of 8736 crystal blocks, 16.1 radiation lengths ($X_0$) thick.
Their energy resolution is 1.8\% for photons above 3 GeV.
The ECL covers the same angular region as the CDC.
Electron identification~\cite{EID} in Belle is based on a combination of
\dedx measurements in the CDC, the response of the ACC, the position and
the shape of the electromagnetic shower, as well as the ratio of the
cluster energy to the particle momentum.
The electron identification efficiency is determined from the two-photon
$e^+e^-\rightarrow e^+e^-e^+e^-$  processes to be more than 90\%
for $p >1.0$~GeV/$c$.
The hadron misidentification probability, determined using tagged pions from
inclusive $K_S^0\rightarrow \pi^+\pi^-$ decays, is below $0.5\%$. 

All the detectors mentioned above are inside a super-conducting
solenoid of 1.7~m radius that generates a 1.5~T magnetic field.
The outermost spectrometer subsystem is a $K_L^0$ and muon detector
(KLM)\cite{KLM}, which consists of 14 layers of iron absorber
(4.7~cm thick) alternating with resistive plate counters (RPC).
The KLM system covers polar angles between 20 and 155 degrees.
Muon identification is based on the depth of penetrated KLM layers and the
position matching from the CDC.
The overall muon identification efficiency, determined by using the two-photon
process $e^+e^-\rightarrow e^+e^-\mu^+\mu^-$  and simulated muons embedded
in $B\overline{B}$ candidate events, is greater than 90\% for tracks with
$p > 1$~GeV/{\it c} detected in the CDC.
The corresponding pion misidentification probability, determined using
$K_S^0\rightarrow \pi^+\pi^-$  decays, is less than 2\%.

\section{Flavor Tagging Algorithm}
\label{sec:algorithm}

\subsection{Principle of Flavor Tagging}
\label{sec:principle}
We determine the flavor of the accompanying $B$ meson based on the flavor
information of the final state particles that belong to $f_{\rm tag}$.
Namely, the flavor of $f_{\rm tag}$ can be determined from the charge
(flavor) of
\begin{enumerate}
\item high-momentum leptons from $B^0 \rightarrow
X \ell^+ \nu$ decays,
\item kaons, since the majority of them originate from
$B^0 \rightarrow K^+ X$ decays through the cascade transition
$\overline b \rightarrow \overline c \rightarrow \overline s$,
\item intermediate momentum leptons from
$\bar{b} \rightarrow \bar{c} \rightarrow \bar{s} \ell^- \overline{\nu}$ decays,
\item high momentum pions coming from $B^0 \rightarrow D^{(*)}\pi^+ X$ decays,
\item slow pions from
$B^0\rightarrow D^{*-}X, D^{*-} \rightarrow \overline D{}^0 \pi^-$ decays, and
\item $\overline{\Lambda}$ baryons from the cascade decay
$\bar{b} \rightarrow \bar{c} \rightarrow \bar{s}$.
\end{enumerate}

The flavor tagging algorithm cannot always determine the flavor of the $B$
mesons from the final state particles; {\it i.e.} $\epsilon < 1$ in general.
This is caused by:
\begin{itemize}
\item inefficiency in particle detection and identification,
\item flavor-nonspecific decay processes such as
$\overline D{}^0 \pi^0, \overline D{}^0 \rightarrow K^0 \pi^0$,
\item processes that have very little information on $b$-flavor such as
$b \rightarrow c \overline{u}d, c \rightarrow \overline K{}^0 X$,
for which the charged particles in the final state are all pions.
\end{itemize}
The incorrect assignment of the flavor is mainly caused by
\begin{itemize}
\item particle misidentification and
\item smaller physical processes that give a flavor estimate that is
opposite to the dominant process, e.g. the charged kaon from the
$\overline{c}$ decay in $b\rightarrow c \overline{c} s$ processes.
\end{itemize}

To maximize $\epsilon_{\rm eff}$, we need to maximize $\epsilon$, and
minimize $w$ using all available information for each event.
As described in detail in the following subsections, a larger
$\epsilon_{\rm eff}$ is obtained by treating events with large $w$'s
and small $w$'s separately.
For this purpose, we use an expected event-by-event dilution factor $r$,
which is described in detail in the next section.
We first find a signature of the aforementioned flavor specific categories
in each charged track and/or a $\Lambda$ baryon candidate in an event.
We assign $r$ to each track/$\Lambda$ candidate.
We combine all such particle level $r$'s taking their correlations
into account, and estimate the $r$ value for the event.
We classify events into six regions according to their $r$ values.
The $r$ value is determined using Monte Carlo (MC) simulation and is related
to $w$ as $r = 1 - 2w$ if the MC simulates the data perfectly.
For each region, we assign a wrong tag fraction $w$, which is measured using
the control data sample.
The values of $w$ are used along with decay time information in an unbinned
maximum likelihood fit to determine the asymmetry parameter $\sin 2 \phi_1$.

\subsection{Flavor Tagging Algorithm}
\label{sec:subalgorithm}

We use two parameters, $q$ and $r$, as the flavor tagging outputs.
The parameter $q$ is the flavor of the tag-side $B$,
as defined in Section~\ref{sec:intro}.
The parameter $r$ is an expected flavor dilution factor that ranges from
zero for no flavor information $(w \simeq 0.5)$ to unity for unambiguous
flavor assignment $(w \simeq 0)$.
In order to obtain a high overall effective efficiency,
we must assign the best estimated flavor dilution factor to each event.
To best accomplish this, we use multiple discriminants in the event.
Using a multi-dimensional look-up table binned by the values of
the discriminants, the signed probability, $q\cdot r$, is given by
\begin{equation}
q \cdot r = \frac{N(B^0) - N(\bzb)}{N(B^0) + N(\bzb)} ,
\label{qr}
\end{equation}
where $N(B^0)$ and $N(\bzb)$ are the numbers of $B^0$ and $\bzb$
in each bin of the look-up table prepared from a large statistics
MC event sample. For example, consider a table with only one bin.
The tagging efficiency and effective tagging efficiency can be written as
\begin{eqnarray}
  \epsilon & \propto & N_0(B^0) + N_0(\bzb), \\
  \epsilon r^2 & = & \epsilon |q \cdot r|^2 \propto
  \frac{[N_0(B^0) - N_0(\bzb)]^2}{N_0(B^0) + N_0(\bzb)}.
\end{eqnarray}
If we subdivide this bin into two bins, the tagging efficiencies,
flavor dilution factors and effective tagging efficiencies can be written as
\begin{eqnarray}
  (q \cdot r)_i & = & \frac{N_i(B^0) - N_i(\bzb)}{N_i(B^0) + N_i(\bzb)}, \\
  \epsilon_i & \propto & N_i(B^0) + N_i(\bzb), \\
  \epsilon_i r_i^2 & = & \epsilon_i |q \cdot r|_i^2
  \propto \frac{[ N_i(B^0) - N_i(\bzb) ]^2}{N_i(B^0) + N_i(\bzb)}.
\end{eqnarray}
Using $\sum_{i=1,2}N_i(B^0) = N_0(B^0)$ and
$\sum_{i=1,2}N_i(\bzb) = N_0(\bzb)$, the sum of the
effective efficiencies of the two bins becomes
\begin{equation}
\sum_{i=1,2} \epsilon_i r_i^2  =  \epsilon r^2 +
\frac{\epsilon_1\epsilon_2(r_1-r_2)^2}{\epsilon} \ge \epsilon r^2 .
\end{equation}
If $r_1 \neq r_2$, then the effective efficiency increases by subdividing.
This can easily be generalized to the case of subdivision into $n$ bins.
The increase of the effective tagging efficiency is proportional
to the dispersion of the flavor dilution factor, $\overline{(r - \bar{r})^2}$
and is always positive.
The number of of bins is practically limited by the quality and quantity
of the Monte Carlo simulation data sample.
Therefore, bins with a large dispersion of $r$
and with sufficient Monte Carlo statistics are subdivided.

\begin{figure}[p]
\begin{center}
\resizebox{0.8\textwidth}{!}{\includegraphics{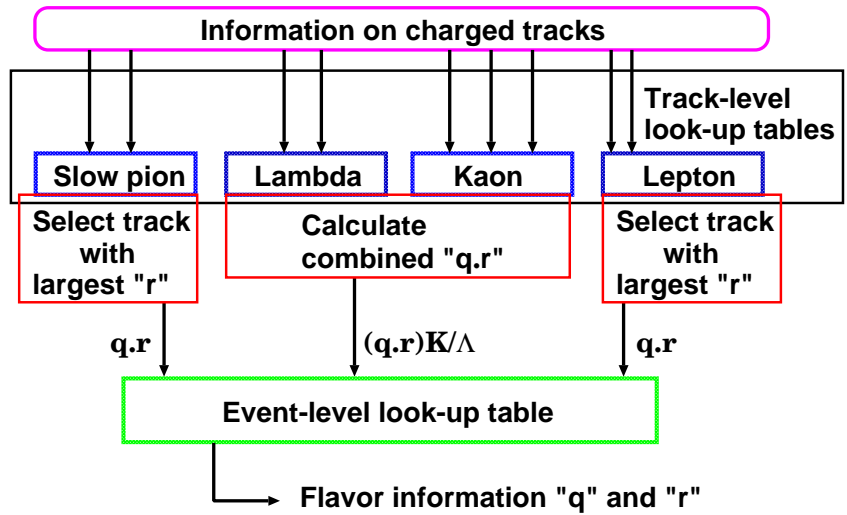}}
\vspace{1ex}
\caption{A schematic diagram of the two-stage flavor tagging.
See the text for the definition of the parameters ``$q$" and ``$r$".}
\label{total}
\end{center}
\end{figure}

Figure \ref{total} shows a schematic diagram of the flavor tagging method.
The flavor tagging proceeds in two stages: the track stage and the event stage.
In the track stage, each pair of oppositely charged tracks is
examined to satisfy criteria for the $\Lambda$-like particle category.
The remaining charged tracks are sorted into slow-pion-like,
lepton-like and kaon-like particle categories.
The $b$-flavor and its dilution factor of each particle,
$(q \cdot r)_X$, in the four categories is estimated using discriminants such
as track momentum, angle and particle identification information.
In the second stage, the results from the first stage
are combined to obtain the event-level value of $q \cdot r$.

In the Belle detector, there is a small asymmetry between particle
and anti-particle production and detection.
For example, the observed yields and signal-to-noise ratios of $\Lambda$
and $\bar{\Lambda}$ candidates are different due to differences in
interactions of the protons and anti-protons in the detector and
differences in their yields in the background.
In our method, $\Lambda$ and $\bar{\Lambda}$ are treated separately and the
effect of the small charge asymmetry is automatically taken into account.
The ${\Lambda}$ candidates have higher tagging efficiency than $\bar{\Lambda}$
candidates due to the larger yields of protons in the background.
On the other hand, $r$ for $\Lambda$'s is lower than $r$ for $\bar{\Lambda}$'s
as the look-up table for $\Lambda$'s contains larger backgrounds
that are generated correctly in the MC simulation.
For other tagging categories such as lepton-, kaon- and slow-pion-like tracks,
there are also small asymmetries,
which are treated in the same way as ${\Lambda}$ candidates.

Using the MC-determined flavor dilution factor $r$ as a measure of the
tagging quality is a straightforward and powerful way of taking into
account correlations among various tagging discriminants.
Using two stages, we keep the look-up tables small enough to provide
sufficient MC statistics for each bin.
In the following we provide details about each stage of the flavor tagging.
Four million $B^0 \bzb$ MC events corresponding to eight million $B$'s
are used to generate the particle-level look-up tables.
To reduce statistical fluctuations of the $r$ values in the particle-level
look-up tables, the $r$ value in each bin is calculated by including events
in nearby bins with small weights.
The event-level look-up table is prepared using MC samples that are
statistically independent of those used to generate the track-level tables
to avoid any bias from a statistical correlation between the two stages.
Seven million $B^0\bzb$ MC events corresponding to fourteen million $B$'s
are used to create the event-level look-up table.
We use GEANT3\cite{geant3} to fully simulate the detector.
Two event generators, QQ\cite{QQ} and EvtGen\cite{EvtGen},
are used to simulate the tag-side $B$ meson decays.
We used QQ-generated MC for early measurements of
$\sin 2 \phi_1$~\cite{bib:cpv,bib:cpv_prd} and EvtGen-generated MC
for more recent $\sin 2 \phi_1$ measurements~\cite{bib:cpv_update}.

\subsection{Particle-level Flavor Tagging}
\label{sec:tracklevel}

For the particle-level flavor tagging, charged tracks that do not belong to
$f_{CP}$ and that satisfy the impact parameter requirements $|dr| < 2$~cm
and $|dz| < 10$~cm are considered.
To find $K_S^0$ and $\Lambda$ candidates, we also use pairs of oppositely
charged tracks that do not belong to $f_{CP}$ according to
a secondary vertex reconstruction algorithm.
Tracks that are a part of a $K_S^0$ candidate or a $\Lambda$ candidate
are not used.
However, the number of $K_S^0$'s in the event is used as a discriminant
in the $\Lambda$-like and kaon-like particle categories.

\subsubsection{Electron-like and Muon-like Track Categories}
\label{sec:lepton}

A track is assigned to the electron-like track category 
if the CMS momentum $p^{\rm cms}_\ell$ is larger than 0.4 GeV/$c$ and
the ratio of its electron and kaon likelihoods is larger than 0.8.
A track is passed to the muon-like track category
if the track has $p^{\rm cms}_\ell$ larger than 0.8 GeV/$c$ and the
ratio of its muon and kaon likelihoods is larger than 0.95.
The likelihoods are calculated by combining the ACC, TOF, \dedx, and
ECL or KLM information.

The discriminants for lepton-like track categories are summarized in
Table~\ref{table:lepton}.
The charge of the particle provides the $b$-flavor $q$, and
other discriminants determine its quality $r$.
The identifier ``$e$ or $\mu$'' specifies whether a track belongs to the
electron-like or the muon-like track categories.
The lepton identification is optimized to reduce the kaon contamination,
while a substantial fraction of pions is included in the muon-like track
category.
Prompt pions from the virtual $W$ decay sometimes preserve the
charge (flavor) of the $W^{\pm}$ boson and therefore can be used to
identify the flavor of the $B$ meson.
About a half of such pions are included in the muon-like track categories
and the rest, in the kaon-like track category.
Such pions in the muon-like track category are included in the bins with
low lepton ID probability and with high momentum.

The variables $p^{\rm cms}_\ell$ and $\theta_{\rm lab}$ are used to
subdivide the table as the purity and the efficiency
of the lepton identification vary as a function of these variables.
The $p^{\rm cms}_\ell$ requirement is also set to accept most of
the primary leptons from the $B$ decays but includes some contamination
due to secondary leptons from $D$ decays.
Since the leptons from cascade $D$ decays give an incorrect flavor
assignment, separation from the primary leptons in $B$ decay is important.
The variable $p^{\rm cms}_\ell$ discriminates primary leptons that tend to
have higher momenta.
The variables $M_{\rm recoil}$ and $P^{\rm cms}_{\rm miss}$ are calculated
using all the observed charged and neutral particles that do not belong to
$f_{CP}$.
A neutrino from semileptonic $B$ decay carries away more momentum than one
from a semileptonic $D$ decays.
The $M_{\rm recoil}$ distribution for semileptonic $B$ decays peaks
around the $D$ mass and has a tail toward the lower side due to
missing particles, while the one for semileptonic $D$ decay distributes
widely up to 5 ${\rm GeV}/c^2$ since $M_{\rm recoil}$ is calculated
including the decay products of the primary $B$ meson.

\begin{figure}[p]
\begin{center}
\resizebox{0.48\textwidth}{!}{\includegraphics{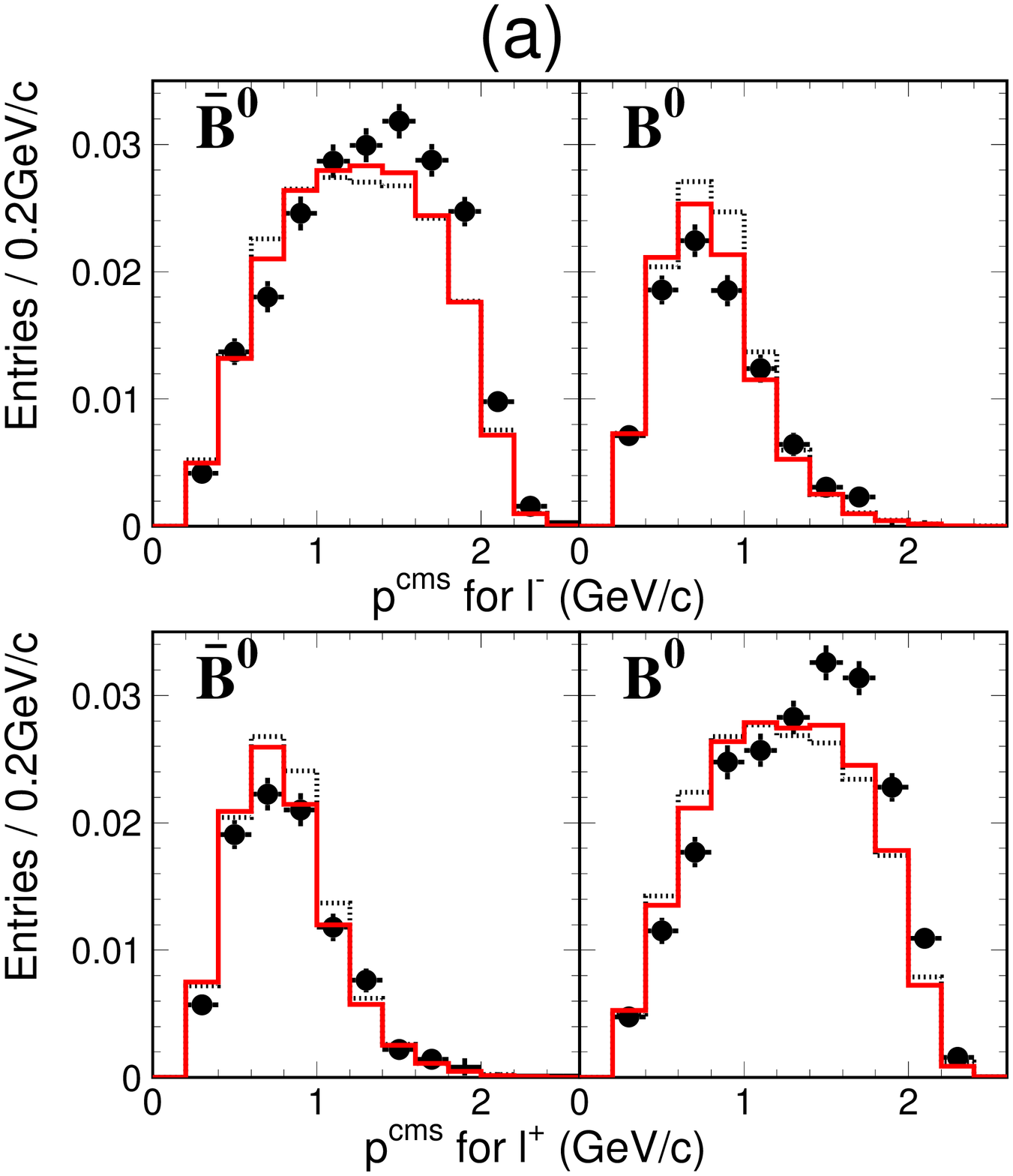}}
\resizebox{0.48\textwidth}{!}{\includegraphics{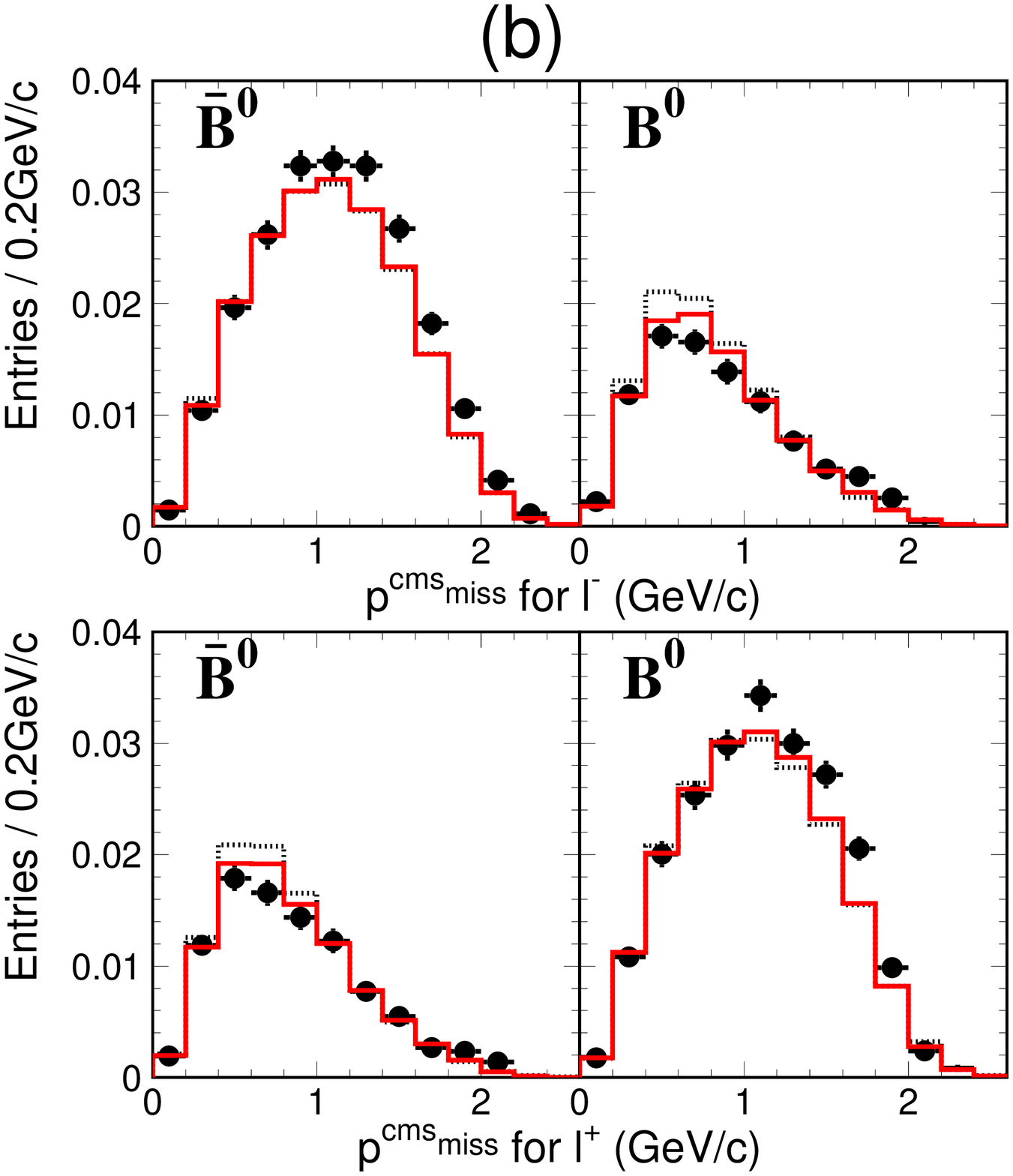}}
\resizebox{0.48\textwidth}{!}{\includegraphics{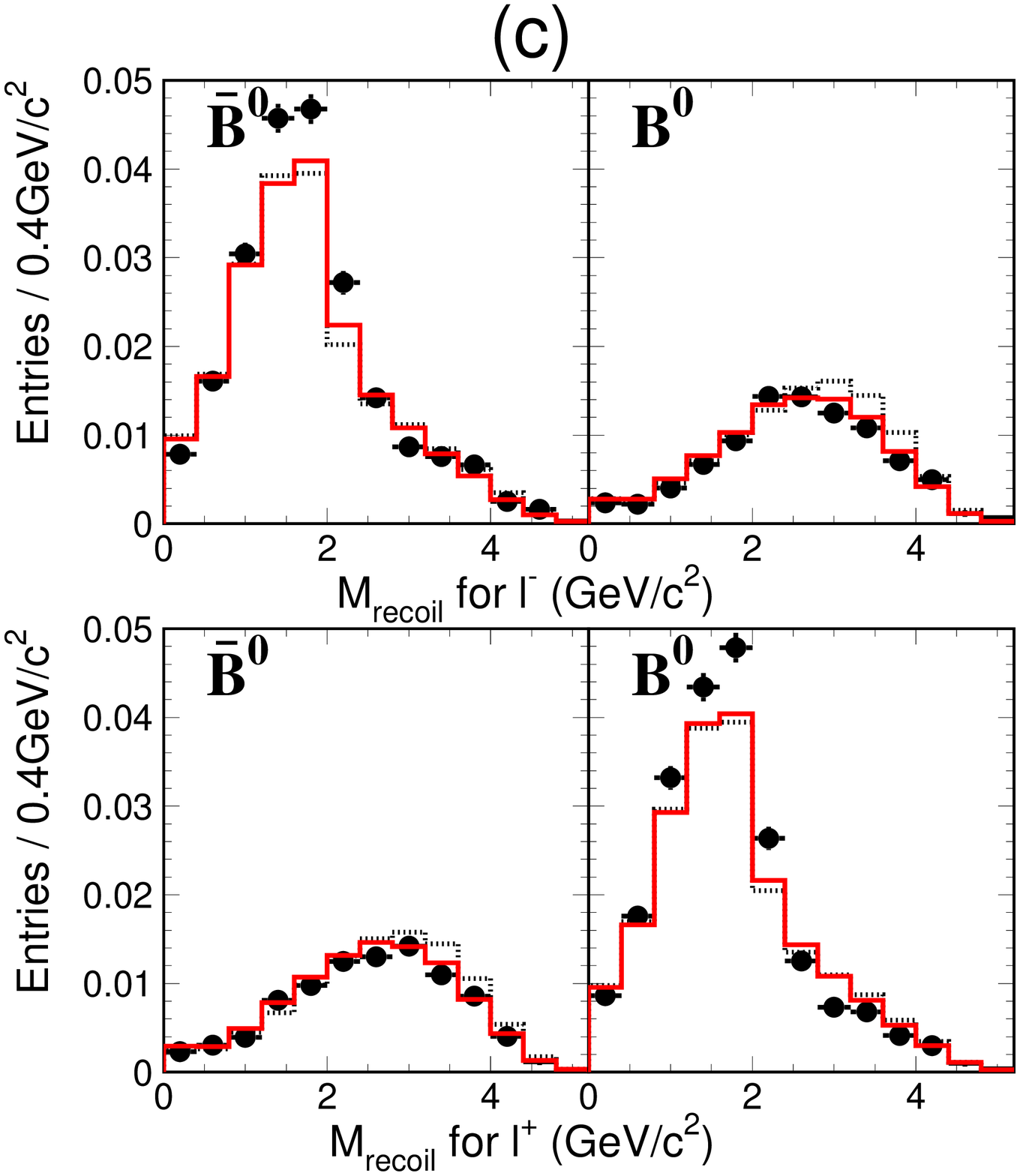}}
\vspace{1ex}
\caption{ (a) $p^{\rm cms}_\ell$, (b) $P^{\rm cms}_{\rm miss}$
  and (c) $M_{\rm recoil}$ distributions for $\bzb$ and $B^0$.
  The points with error bars are control sample data
  (See Appendix~\ref{sec:ctrl_samples}).
  The solid and dotted histograms are the EvtGen-MC and QQ-MC, respectively.
  All distributions are made with a requirement on lepton ID in
  Table~\ref{table:lepton} to remove the large pion background.
  The upper two figures and lower two figures in (a), (b) or (c)
  are for $\ell^-$-like tracks and for $\ell^+$-like tracks, respectively.
  The upper left and lower right figures in (a), (b) or (c)
  contain primary leptons from $B$ decay,
  while upper right and lower left figures in (a), (b) or (c)
  contain secondary leptons from $D$ decay.
}
\label{fig:lepton}
\end{center}
\end{figure}

Figure~\ref{fig:lepton} shows the $p^{\rm cms}_\ell$, $M_{\rm recoil}$
and $P^{\rm cms}_{\rm miss}$ distributions for the data and the MC.
Although some disagreement is visible, the experimental bias due to this
disagreement is found to be negligible since $w$ is evaluated from
control samples as described in Section~\ref{sec:performance}.

Within the lepton categories, leptons from semileptonic $B$ decays
yield the highest effective efficiency while leptons from $B \rightarrow D$
cascade decays and high-momentum pions from $B^0\rightarrow D^{(*)-}\pi^+X$
make small additional contributions.

\subsubsection{Slow-pion-like Track Category}
\label{sec:slowpi}

A track that has CMS momentum below 0.25 GeV/$c$ and is not identified
as a kaon, is assigned to the slow-pion-like track category.
The discriminant variables for the slow-pion-like track category are given in
Table \ref{table:slowpi}.
The largest background is from other ({\it i.e.} non-$D^*$ daughter)
low momentum pions.
Since the $Q$ value is small,
the pion from the $D^{*+} \rightarrow D \pi^+$ decay has a low momentum
and has a flight direction that follows the $D^*$ direction.
We use $\alpha_{\rm thr}$, the angle between the direction of the
slow-pion-like track and the axis of the thrust calculated from the tag-side
particles in the CMS to select $B\rightarrow D^{*-}\pi, \rho$ decays.

\begin{figure}[p]
\begin{center}
\resizebox{0.48\textwidth}{!}{\includegraphics{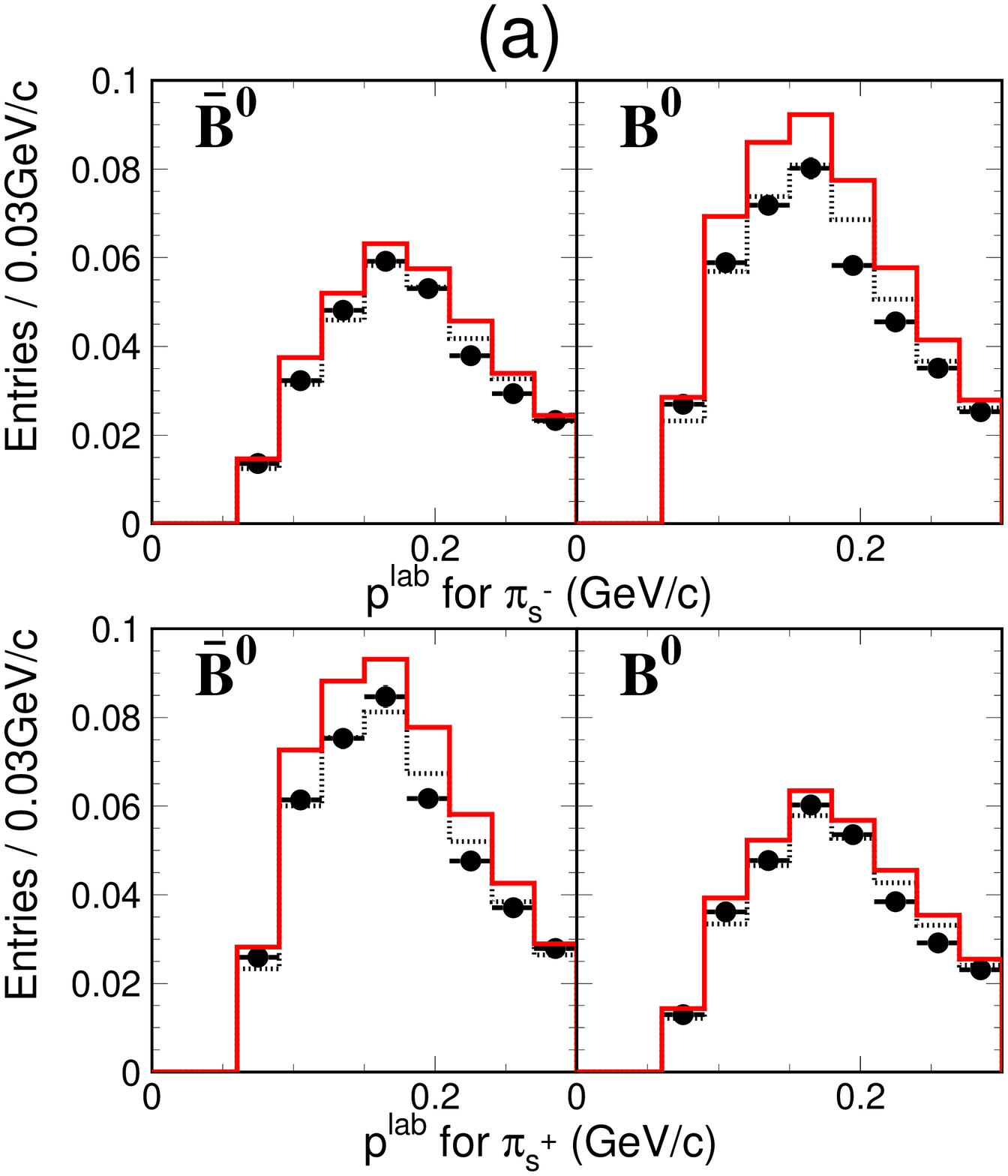}}
\resizebox{0.48\textwidth}{!}{\includegraphics{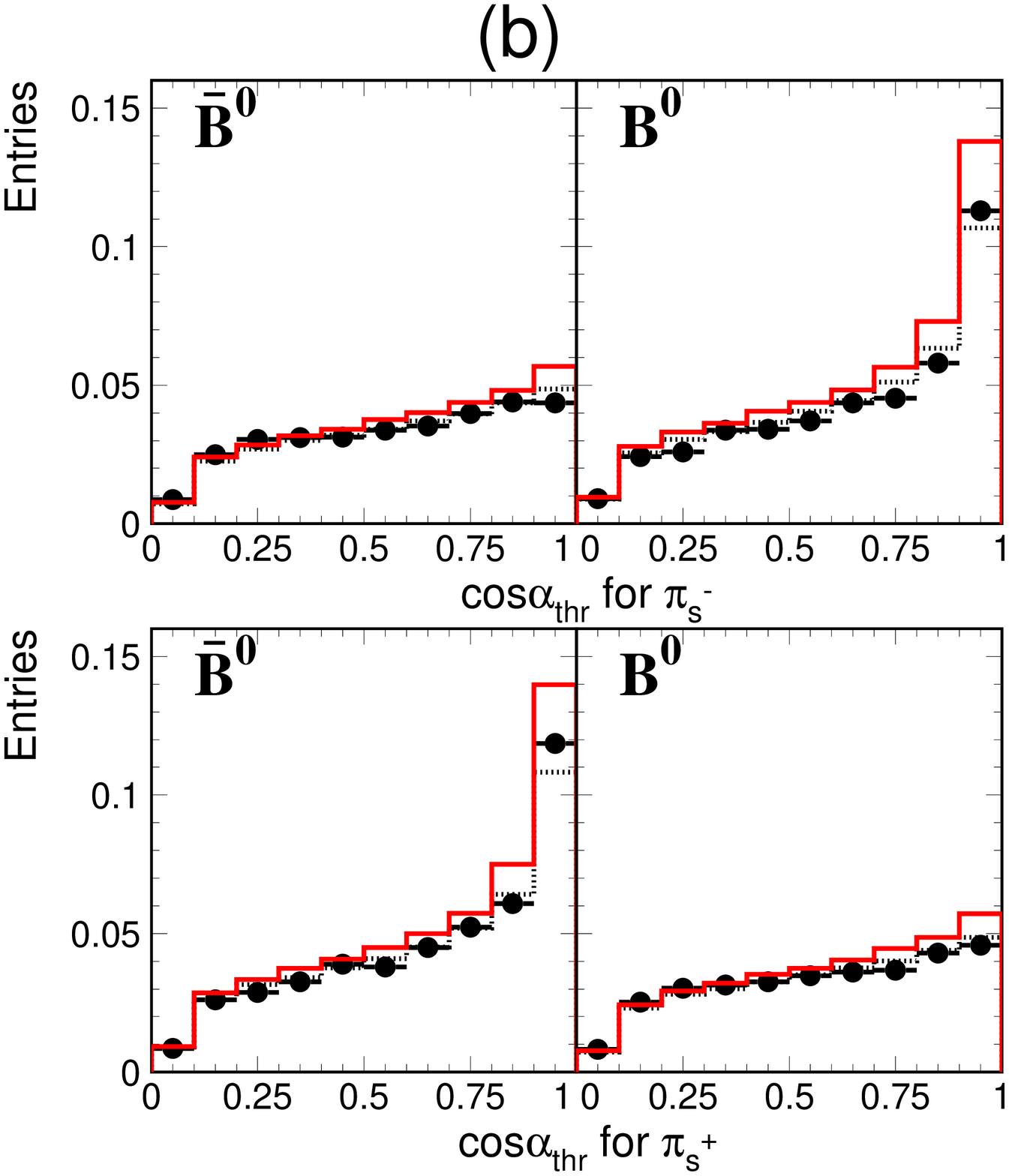}}
\vspace{1ex}
\caption{ (a) $p^{\rm lab}$ and (b)
 $\cos\alpha_{\rm thr}$ distributions of slow pion for $\bzb$ and $B^0$.
 The points with error bars are control sample data
 (See Appendix~\ref{sec:ctrl_samples}),
 while the solid and dotted histograms are the EvtGen-MC and QQ-MC,
 respectively.
 All distributions are made with a requirement on $\pi /e$ ID to remove
 low momentum electrons from photon conversions and $\pi^0$ Dalitz decays.
 The upper two figures and lower two figures in (a) or (b)
 are for $\pi^-_s$-like tracks and for $\pi^+_s$-like tracks, respectively.
 The upper right and lower left figures in (a) or (b)
 contain slow pions from $D^{* \pm}$ decays.
}
\label{fig:slowpi}
\end{center}
\end{figure}

The other background in this category is from electrons produced
in photon conversions and $\pi^0$ Dalitz decays.
Electrons coming from photon conversion are identified through
the secondary vertex reconstruction algorithm and are rejected.
To separate slow pions from the remaining electrons, we use only \dedx{}
because the tracks do not have enough transverse momenta
($p_t > 0.3~{\rm GeV/}c$) to reach the ECL detector,
which gives $E/p$ and other useful information to discriminate electrons
from pions.
The $\pi /e$ ID probability from \dedx{} strongly depends on the
laboratory momentum;
the \dedx{} losses are equal for electrons and pions around
$p_{\rm lab} = 0.2{\rm GeV/}c$.
Thus, we use $p_{\rm lab}$ instead of $p_{\rm cms}$ in this category.
Figure~\ref{fig:slowpi} shows the distribution of $\cos\alpha_{\rm thr}$
and the momenta of the slow pion candidates in the laboratory frame,
which are uniquely determined by the values of $p_{\rm lab}$ and
$\theta_{\rm lab}$.

\subsubsection{Kaon-like Track Category}
\label{sec:kaon}

If a track does not fall into any of the categories described above, and
is not positively identified as a proton, it is classified as a
kaon-like track.
Charged kaons from $b\rightarrow c\rightarrow s$ are included into this
category.
Some pions are also included in this category.
The discriminants for this category are listed in Table~\ref{table:kaon}.
The variables $p^{\rm cms}$, $\theta_{\rm lab}$ and $K/\pi$ ID
separate kaons from pions. For $K/\pi$ identification, the
information from the \dedx{}, TOF and ACC detectors is combined
into a likelihood variable and a single kaon probability is calculated.
The table is subdivided into $p^{\rm cms}$ and $\theta_{\rm lab}$ bins
as the purity of kaon changes as a function of these variables.

The kaon-like track category is subdivided into two parts:
events with and without $K_S^0$ decays (a switch ``w/ or w/o $K_S^0$''),
since they have different purities;
a kaon accompanied with $K^0_S$'s tends to originate from a strange quark
in a $b\rightarrow c\bar{c}(d,s)$ decay or from $s\bar{s}$ popping,
while one without $K^0_S$'s has a higher probability to be
from the cascade decay ($b\rightarrow c \rightarrow s$).

\begin{figure}[p]
\begin{center}
\resizebox{0.48\textwidth}{!}{\includegraphics{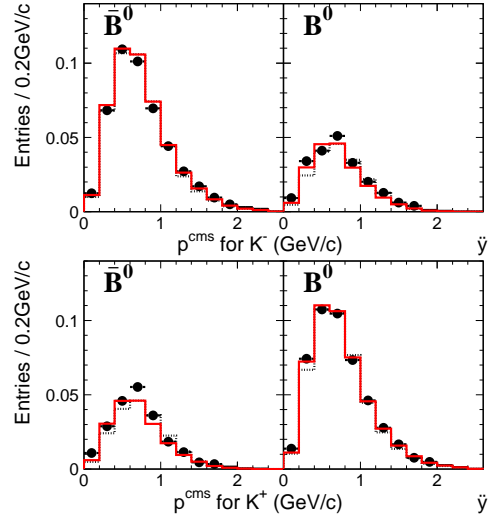}}
\vspace{1ex}
\caption{$p^{\rm cms}$ distributions of kaons for $\bzb$ and $B^0$.
 The points with error bars are control sample data
 (See Appendix~\ref{sec:ctrl_samples}),
 while the solid and dotted histograms are the EvtGen-MC and QQ-MC,
 respectively.
 $K/\pi$ ID in table~\ref{table:kaon} is required to exclude the
 dominating pion background.
 The upper two figures and lower two figures
 are for $K^-$-like tracks and for $K^+$-like tracks, respectively.
 The upper right and lower left figures
 contain kaons from cascade $b\rightarrow c \rightarrow s$ transition.
}
\label{fig:kaonfig}
\end{center}
\end{figure}

The bins for low kaon probability and high $p^{\rm cms}$
contain fast pions. Fast pions from a prompt $B$ decay,
such as $B \rightarrow D \pi(\rho)$, have flavor information and
give some contribution to $\epsilon_{\rm eff}$.
Figure~\ref{fig:kaonfig} shows the CMS momentum distribution of kaon-like
track candidates compared to those in MC.

\subsubsection{$\Lambda$-like Particle Category}
\label{sec:lambda}

\begin{figure}[p]
\begin{center}
\resizebox{0.48\textwidth}{!}{\includegraphics{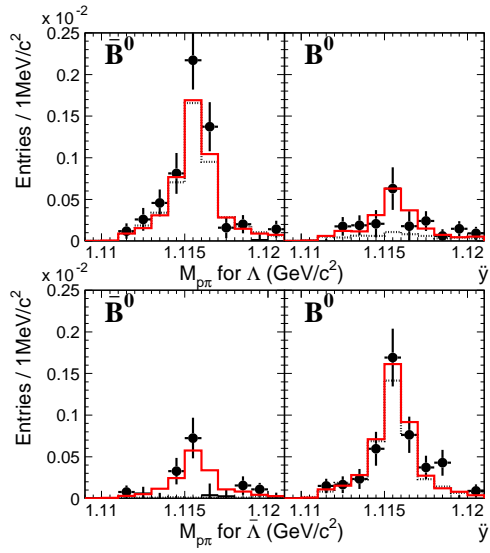}}
\vspace{1ex}
\caption{
 $M_{p\pi}$ distributions of $\Lambda$ candidates for $\bzb$ and $B^0$.
 The points with error bars are the control sample data
 (see Appendix~\ref{sec:ctrl_samples}).
 The solid and dotted histograms are the EvtGen-MC and QQ-MC samples,
 respectively.
 The upper two figures and lower two figures are for $\Lambda$ candidates
 and for $\overline{\Lambda}$ candidates, respectively.
 The upper left figure and lower right figure contain
 $\Lambda$ particles from cascade $b \rightarrow c \rightarrow s$ transition.
}
\label{fig:lambda}
\end{center}
\end{figure}

$\Lambda$ candidates are selected from pairs of oppositely-charged tracks
one of which is identified as proton, and that also satisfy
$1.1108~{\rm GeV}/c^2 < M_{p \pi} < 1.1208~{\rm GeV}/c^2$,
$\theta_{defl} < 30^{\circ}$, $|\Delta z| < 4.0$ cm
and with a secondary vertex position in the $r-\phi$ plane above 0.5 cm.
Figure~\ref{fig:lambda} shows the $M_{p\pi}$ distributions
for the data and the MC.
The discriminants for this category as well as the definitions of
$M_{p \pi}$, $\theta_{defl}$ and $\Delta z$ are listed in
Table~\ref{table:lambda}.
The last column is the number of bins for the corresponding discriminant.

As the purity (signal-to-noise ratio) of the $\Lambda$ candidates
varies as a function of the variables $M_{p \pi}$, $\theta_{defl}$,
and $\Delta z$, we subdivide this category using these variables.
Since the number of $\Lambda$ candidates is small, for each
discriminant we subdivide into two (high and low quality) bins.
The high quality bin contains $\Lambda$ candidates with
$1.1148~{\rm GeV}/c^2 < M_{p\pi} < 1.1168~{\rm GeV}/c^2$,
$\theta_{defl}<  10^{\circ}$ and $|\Delta z |<0.5$ cm.

\subsection{Event-level Flavor Tagging}
\label{sec:event}

The track-level $(q\cdot r)_X$'s are combined for event-level tagging.
From the lepton-like and slow-pion-like track categories,
the track with the highest $r$-value from each category is
chosen as the input to the event level look-up table.
The flavor dilution factors of the kaon-like and $\Lambda$-like particle
candidates are combined by calculating the product of the flavor
dilution factors in order to account for the cases with multiple $s$ quark
contents in an event.
The product of flavor dilution factors gives better effective efficiency
than taking the track with the highest $r$.
Table~\ref{table:event} shows the discriminants for event-level tagging.
By using a three-dimensional look-up table,
the correlations between flavor information for lepton-like, slow-pion-like,
kaon-like and $\Lambda$-like particles are correctly taken into account.
In the event-level look-up table, one of the bins in Table~\ref{table:event}
corresponds to ``empty'' for the case
when there is no output from a particular particle-level category.

\begin{figure}[p]
\begin{center}
\resizebox{0.48\textwidth}{!}{\includegraphics{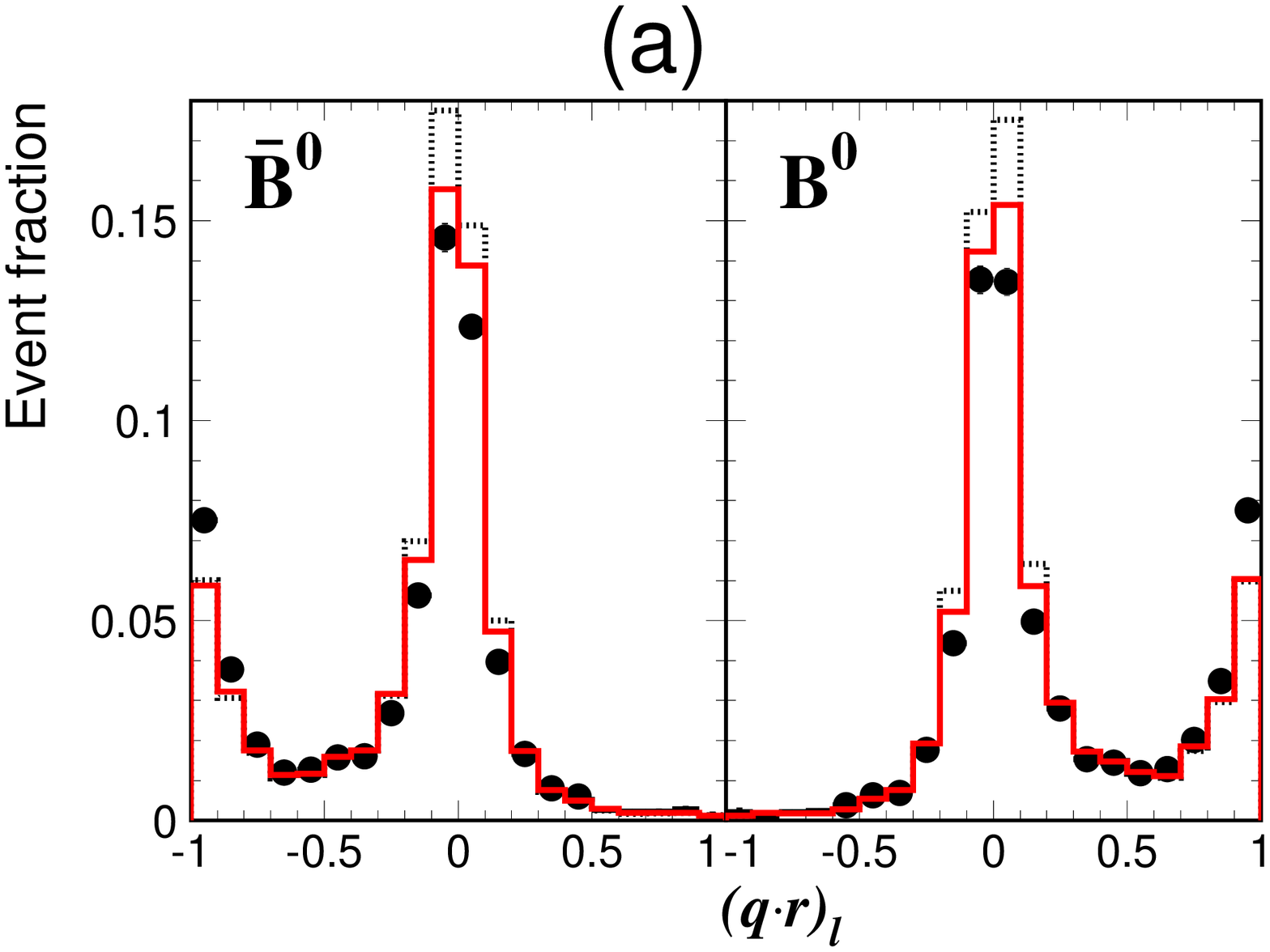}}
\resizebox{0.48\textwidth}{!}{\includegraphics{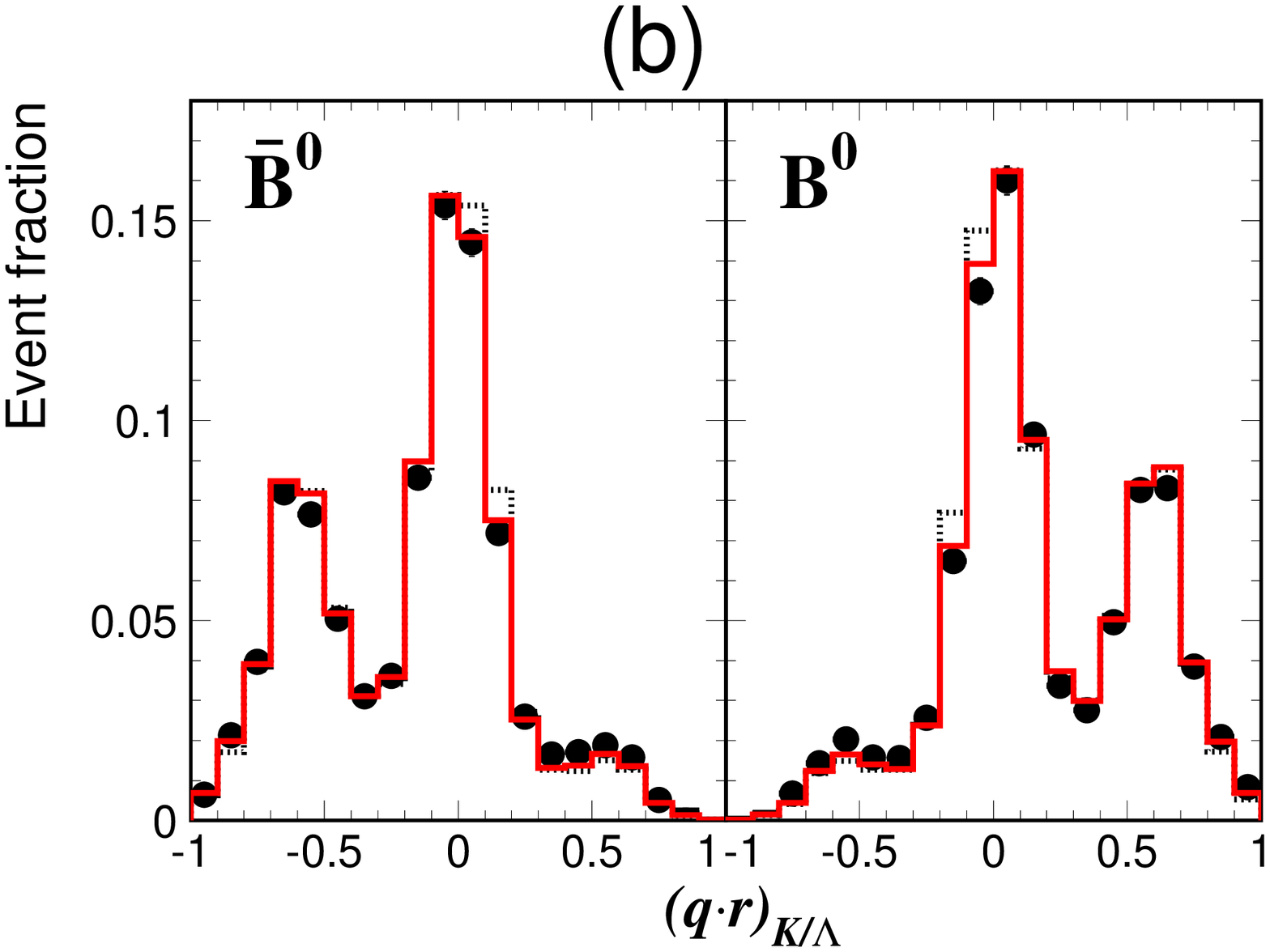}}
\resizebox{0.48\textwidth}{!}{\includegraphics{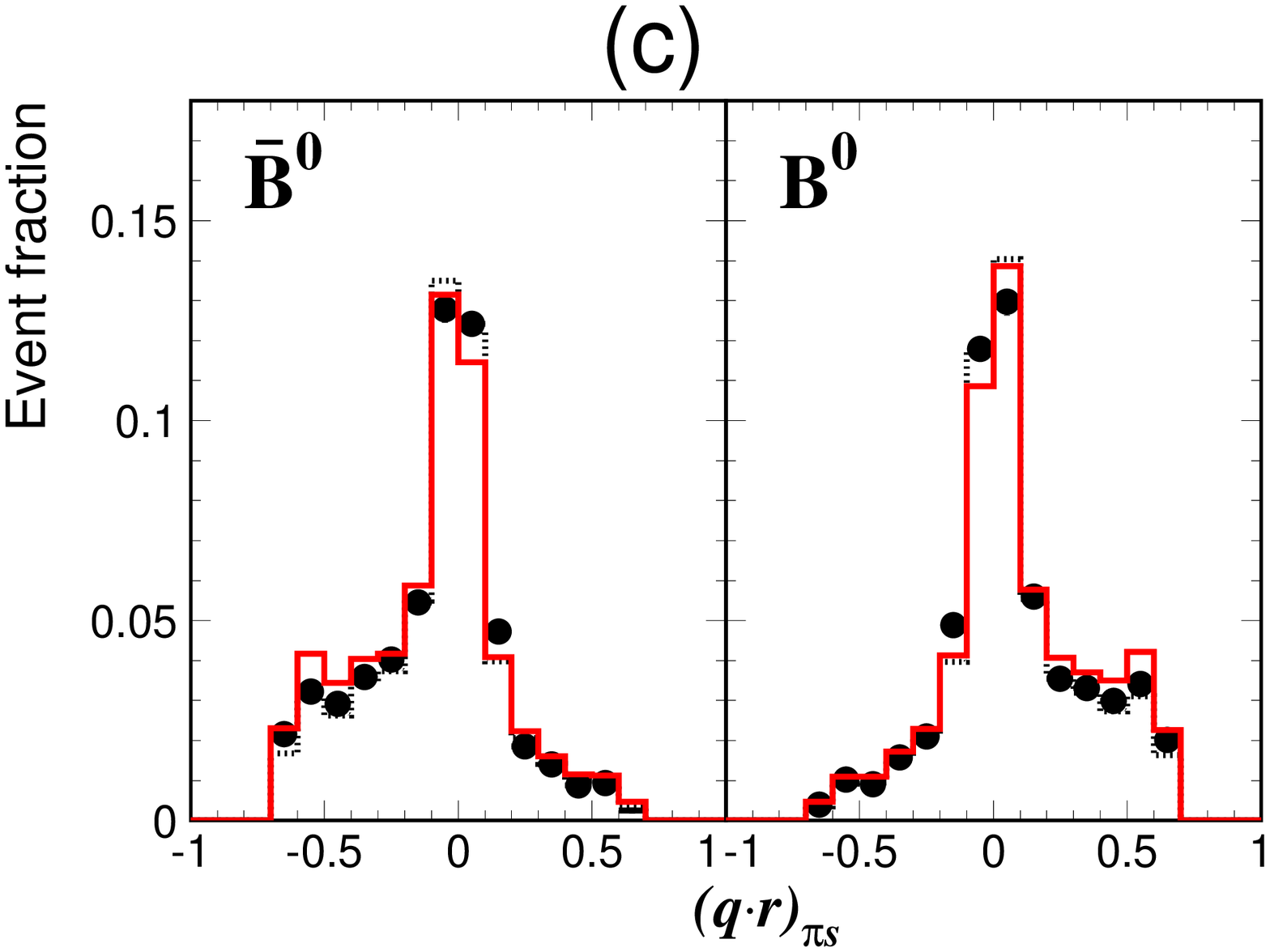}}
\vspace{1ex}
\caption{ Distributions of event layer input variables
  (a) $(q \cdot r)_{l}$, (b) $(q \cdot r)_{K/\Lambda}$ and
  (c) $(q \cdot r)_{\pi_s}$ for $\bzb$ and $B^0$.
  The points with error bars are control sample data
  (see Appendix~\ref{sec:ctrl_samples}).
  The solid and dotted histograms are the EvtGen-MC and QQ-MC, respectively.
  For these distributions, the $(q \cdot r)$ values of corresponding
  track-layer outputs are obtained with the EvtGen-MC lookup table.
  Events with no input tracks that have $r=0$ are excluded.
  The fractions of such events are 36\%, 10\% and 41\% for the lepton,
  the kaon/$\Lambda$ and the slow pion categories, respectively.
}
\label{fig:event}
\end{center}
\end{figure}

Figure~\ref{fig:event} shows the distributions of input values for the
event layer.
The distribution of $(q \cdot r)_{\rm lepton}$ has peaks around $\pm 1$
due to the high momentum primary leptons from semileptonic $B$ decays.
The effective efficiency for the lepton category is 12\%
according to MC simulation.
The distribution of $(q \cdot r)_{K/\Lambda}$ has peaks around $\pm 0.6$,
which correspond to events with a single kaon or a $\Lambda$ candidate.
The entries around $\pm 1.0$ correspond to events with multiple kaon and/or
$\Lambda$ candidates with consistent flavor information.
Kaons have higher yields but less flavor information compared to leptons.
The combined effective efficiency estimated for kaon and $\Lambda$ categories
is 18\% according to MC.
The distribution of $(q \cdot r)_{\pi_s}$ contains no entries beyond $\pm 0.7$,
as the slow-pion-like track category has much more background
than other categories.
The effective efficiency for the slow-pion-like track category is
estimated to be 6\% according to MC.
The peaks around zero in the three distributions are due to
pions that have little flavor information.

\begin{figure}[p]
\begin{center}
\resizebox{0.8\textwidth}{!}{\includegraphics{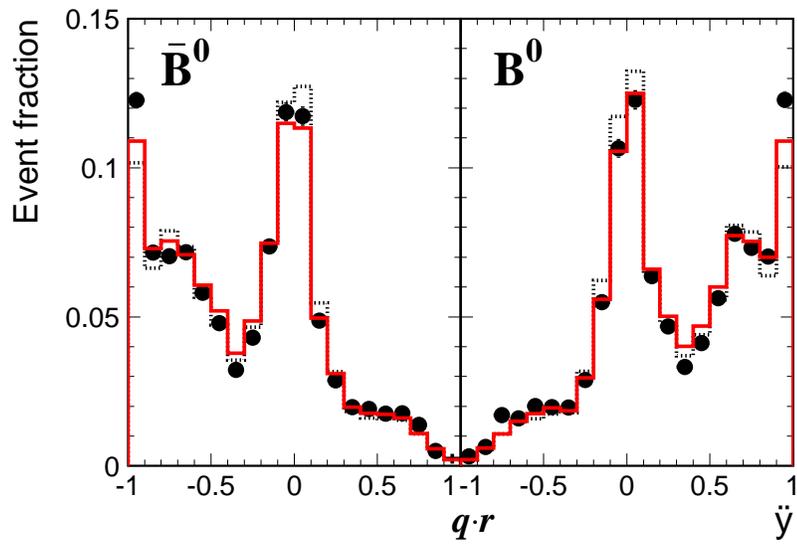}}
\vspace{1ex}
\caption{The $q\cdot r$ distribution for $\bzb$ and $B^0$.
  The points with error bars are control sample data (See Appendix~\ref{sec:ctrl_samples}),
  while the solid and dotted histograms
  are the EvtGen-MC and the QQ-MC, respectively. }
\label{fig:mdlh_data_vs_mc}
\end{center}
\end{figure}

The probability that we can assign a non-zero value for $r$ is 99.6\%
according to MC; {\it i.e.} almost all the reconstructed $f_{CP}$ candidates
can be used to extract $\sin 2 \phi_1$.
Using a MC sample that is statistically independent of those used to generate
the look-up tables,
we estimate the effective efficiency to be $29.3 \pm 0.1 \%$.
Since the lepton-like, kaon/$\Lambda$-like and slow-pion-like particle tagging
categories are not exclusive,
the effective tagging efficiency is smaller than the sum of the efficiencies
for the individual particle categories.
We compare the distribution of $q\cdot r$ obtained from
the control data samples with the MC expectation.
As shown in Figure~\ref{fig:mdlh_data_vs_mc},
the data and MC are in good agreement.

Finally, we obtain the wrong tag fraction, $w$ for each event from the
MC-determined event-level flavor dilution factor, $r$.
Trusting the Monte Carlo simulation completely
we could assign $w$ for each event from the relation, $r = 1 - 2w$.
However, using the method described in the following section,
we measure $w$ using the control data samples of flavor specific $B^0$ decays
and use the measured $w$ values in the unbinned maximum likelihood fit to
extract $\sin2\phi_1$. All tagged events are sorted into six subsamples
according to the value of $r$; $0<r\leq 0.25$, $0.25 < r \leq 0.5$,
$0.5 < r \leq 0.625$, $0.625 < r \leq 0.75$, $0.75 < r \leq 0.875$
and $0.875 < r \leq 1$.
Wrong tag fractions $w_l$ are measured for the six regions.
The average value of $r$ for each region ($r_l$)
and measured wrong tag fraction ($w_l$) should satisfy $r_l \simeq 1-2w_l$,
if the MC that is used for constructing the look-up tables simulates
generic $B$ decays correctly.
Using the measured (and therefore average) $w$ value for the region instead
of a $w$ value calculated for each event from MC,
we introduce no systematic bias into the measurement of $\sin 2 \phi_1$
from the Monte Carlo simulation,
although the effective tagging efficiency degrades.
The degradation from the subdivision into $r$ bins is estimated to be
about $\sim 0.5 \%$, according to a Monte Carlo study.
Nevertheless, in such a categorization based on the $r$ value,
we can achieve a higher effective tagging efficiency
than using the traditional method of treating lepton, kaon and slow pion
tags separately; by using a conventional tagging method with
kaons and high momentum leptons, we obtain an total effective
tagging efficiency of $22.2\pm 0.1\%$.

We have also investigated the dependence of flavor tagging performance on MC.
We prepared two sets of lookup tables, a QQ-generated table and
a EvtGen-generated table.
Comparing the QQ-generated and the EvtGen-generated tables,
we find the EvtGen-generated table has the larger effective tagging efficiency.
As a result, we switched to EvtGen-MC tables starting with the updated
$\sin 2 \phi_1$ analysis~\cite{bib:cpv_update}.
In this section, the performance of the flavor tagging with EvtGen-MC
is discussed and that with QQ-MC is referred to for comparison purposes only.
The performance of the latter is described in~\cite{bib:cpv_prd}.

\section{Flavor Tagging Performance}
\label{sec:performance}

The flavor tagging performance is evaluated using the control samples
of self-tagged $B$-meson decays, which are described in
Appendix~\ref{sec:ctrl_samples}.
The flavor tagging efficiency, $\epsilon$ is measured to be 99.8\%,
which is consistent with the MC expectation.
The wrong tag fraction $w$ is obtained by fitting the time-dependent
$B^0$-$\bzb$ mixing oscillation signal.
The analysis method is similar to the one used in the previous
Belle $B^0$-$\bzb$ mixing analysis~\cite{dslnu_mixing,hadron_mixing}.
The time evolution of neutral $B$-meson pairs with opposite flavor (OF)
or same flavor (SF) is given by:
\begin{eqnarray}
{\mathcal P}_{\rm OF(SF)}(\Delta t) &=& \frac{e^{-|\Delta
t|/\tau_{B^0}}}{4\tau_{B^0}}[ 1 \pm  (1-2w)\cos(\Delta m_d \Delta t)],
\end{eqnarray}
and the OF-SF asymmetry,
\begin{eqnarray}
  A_{\rm mix} \equiv { {\mathcal P}_{\rm OF} - {\mathcal P}_{\rm SF} \over
               {\mathcal P}_{\rm OF} + {\mathcal P}_{\rm SF} }
           = (1 - 2w)\cos(\Delta m_d \Delta t).     
\end{eqnarray}
We obtain the wrong tag fraction using an unbinned maximum likelihood
fit to the reconstructed $\Delta t$ distribution of the SF and OF
events with $\tau_{B^0}$ and $\Delta m_d$ fixed to the world average
values~\cite{PDG2002}.  
The likelihood function $L$ is defined as:
\begin{eqnarray}
L = {\Pi}_i P_{OF}(\Delta t_i) \times {\Pi}_j P_{SF}(\Delta t_j),
\end{eqnarray}
where the index $i(j)$ runs over all selected OF(SF) events.
The function $P_{OF(SF)}$ is the sum of signal probability density function
smeared by the $\Delta t$ resolution and the background component,
written as:
\begin{eqnarray}
P_{OF(SF)} &=& f_{\rm sig}[(1-f_{\rm ol})F_{\rm sig}^{OF(SF)}(\Delta
t)+f_{\rm ol}f_{\rm sig}^{OF(SF)}F_{ol}(\Delta t)]  \\
&&+(1-f_{\rm sig})[(1-f_{\rm ol})F_{\rm bkg}^{OF(SF)}(\Delta t) +
f_{\rm ol}f_{\rm bkg}^{OF(SF)}f_{\rm ol}f_{\rm bkg}^{OF(SF)}F_{\rm ol}(\Delta t)],\nonumber \\
F_{\rm sig}^{OF(SF)} &=& \int {\mathcal P}_{OF(SF)}(\Delta t')
R(\Delta t-\Delta t') d\Delta t', \nonumber
\end{eqnarray}
where $f_{\rm sig}$ and $F_{\rm bkg}$ are the signal fraction and
the background $\Delta t$ shape, respectively,
whose description can be found elsewhere~\cite{dslnu_mixing,hadron_mixing},
and $R$ is the $\Delta t$ resolution function.
A small fraction of events at large $\Delta t$ (outliers) are
represented by a Gaussian of a large width, $F_{\rm ol}$.
The outlier-fraction $f_{\rm ol}$, the width of $F_{\rm ol}$ and the
resolution function $R$ are determined in the $B$ lifetime
analysis~\cite{hadron_lifetime}.

\begin{figure}[p]
\begin{center}
\resizebox{0.8\textwidth}{!}{\includegraphics{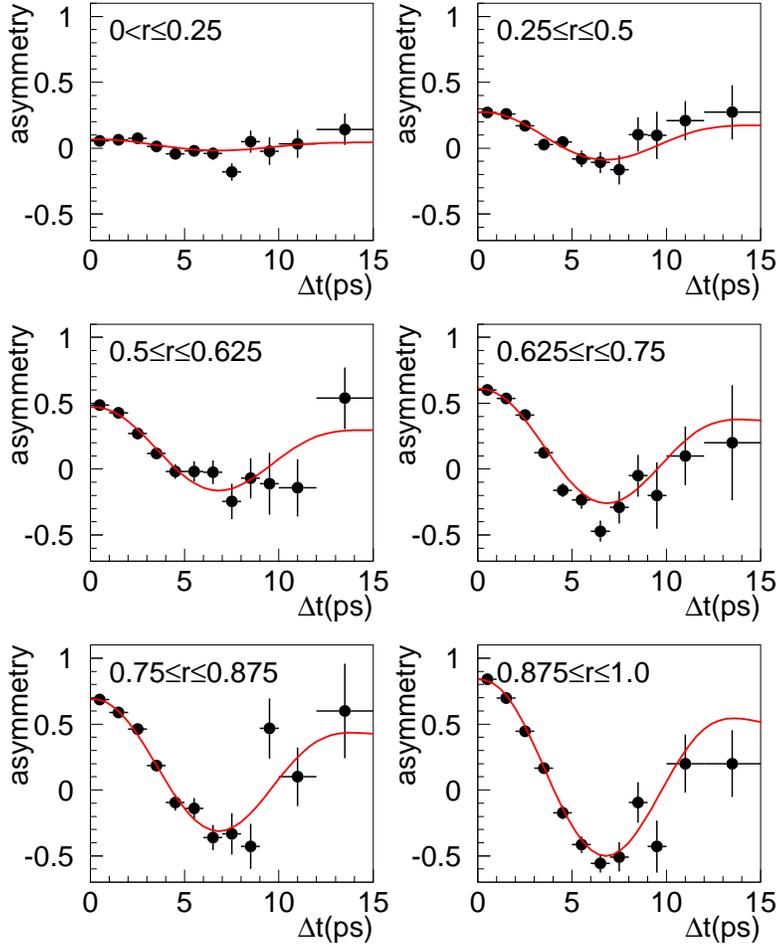}}
\vspace{1ex}
\caption{Measured asymmetries between the OF events and the SF events
         (OF-SF asymmetries)
         for six regions of $r$ obtained from control samples.
	 The definition of OF and SF is given in the text.
	 The background is not subtracted in the asymmetry plots.
         Solid curves show the result of the unbinned maximum likelihood fit.}
\label {fbtg_asym}
\end{center}
\end{figure}

\begin{figure}[p]
\begin{center}
\resizebox{0.6\textwidth}{!}{\includegraphics{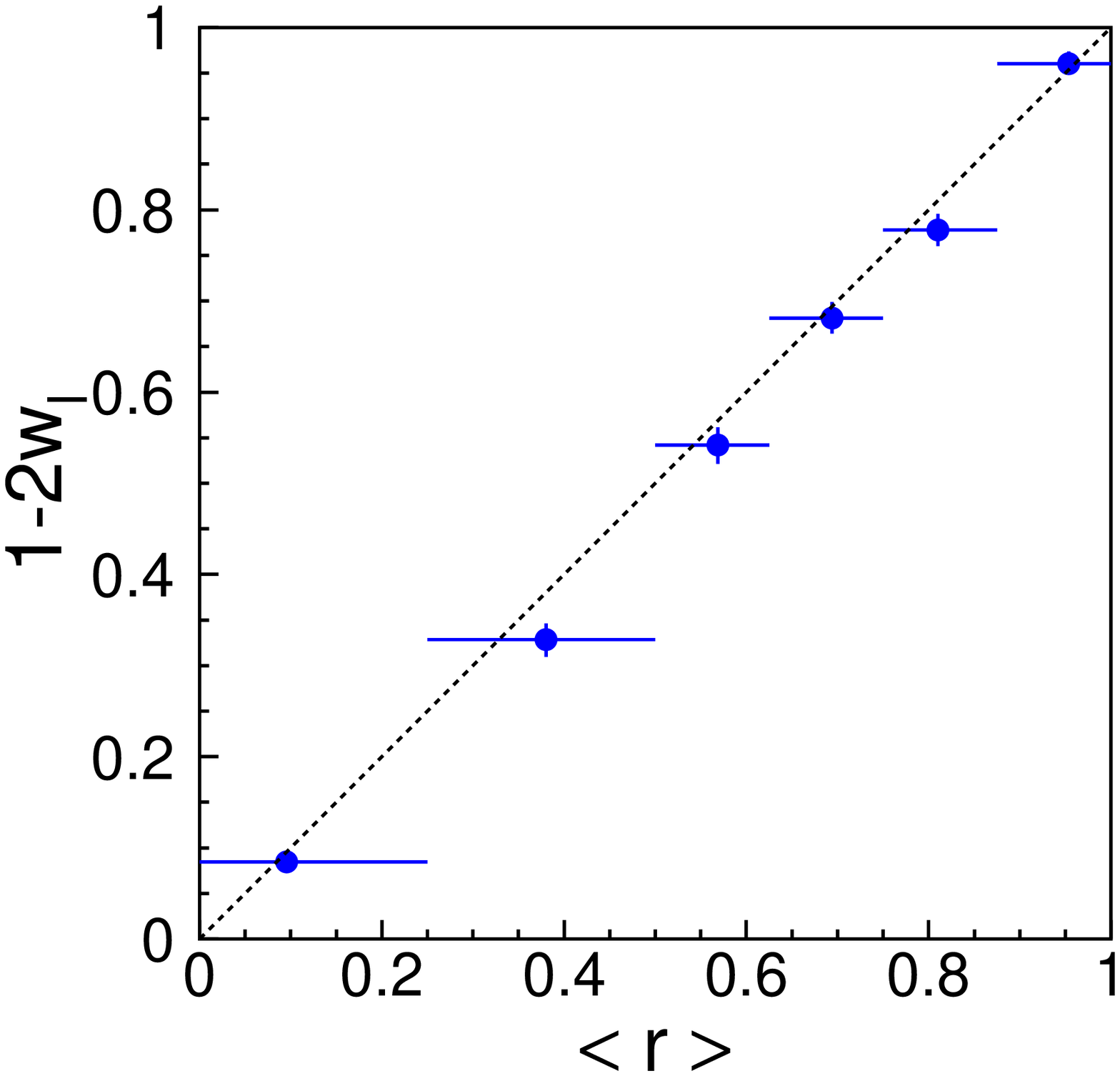}}
\vspace{1ex}
\caption{
  Measured $1-2w$ as a function of the mean value of $r$ in each $r$ region.
  The $<r>$ values are taken from the $\jpsi\ks$ MC.}
\label {fbtg_dilvsr}
\end{center}
\end{figure}

Figure~\ref{fbtg_asym} shows the measured OF-SF asymmetries as a function of
$\Delta t$.
Figure~\ref{fbtg_dilvsr} shows the measured $1-2w$ vs. $r$,
which confirms the validity of our tagging method.

The fit results are summarized in Table~\ref{tag_sum}.
We also evaluate the wrong tag fractions for $B^0$-tagged events and
$\bzb$-tagged samples separately as a check, since they can be different due to
charge asymmetries in the particle identification and proton fakes
in $\Lambda$ candidates.
The $w_l$ values for the two samples are listed in Table~\ref{q_pm_diff}.
We will use separate $w_l$ for $B^0$-tagged and $\bzb$-tagged samples
as the statistics of the control samples increases.
\footnote{
The latest update of $\sin 2 \phi_1$ based on a data sample of
$152 \times 10^6$ $B\overline{B}$ pairs uses separate $w_l$ values for
$B^0$-tagged and $\bzb$ tagged samples.
}

Systematic errors are summarized in Table~\ref{sys_sum}.
The uncertainties in semileptonic modes are the dominant components.
The systematic errors due to the uncertainty of the signal fraction are
estimated by changing each fraction or parameter representing the fraction by
$\pm 1 \sigma$, repeating the fit and adding the deviations from the
main result in quadrature.
For the hadronic modes, we consider the effect of changing the
signal region in the energy difference $\Delta E \equiv E_{B}^{cms}-
E_{\rm beam}$ by $\pm$ 10 MeV and beam-energy constrained mass
$M_{bc}=\sqrt{(E_{\rm beam})^2-(p_{B}^{cms})^2}$ by
$\pm$ 3MeV, where $E_B^{cms}$ and $p_{B}^{cms}$ are the
energy and the momentum of the reconstructed $B$ meson in the $\Upsilon(4S)$
center of mass system.
Each parameter in the $\Delta t$ background shape is varied by $\pm 1 \sigma$,
the fit is repeated and the errors are added in quadrature.
We also check a possible difference between
the $\Delta t$ background shape
in the signal region and the background control sample
using MC simulation and include it in the systematic error.
We use MC to estimate the $B \rightarrow D^{**}\ell\nu$ background
in the semileptonic sample, where $D^{**}$ denotes non-resonant $D^*\pi$ and
heavier charmed mesons.
We evaluate the systematic errors due to
uncertainties of the branching fractions of each $D^{**}$ component by
successively setting each $D^{**}$ component in turn to unity in the MC
(with all others set to zero), and repeating the fit.
We take the largest variation in $w_l$ as the systematic error.
The effect of the uncertainty in the $B^{+} \rightarrow
D^{**-}\ell^{+}\nu$ background for the semileptonic mode is also
included.
Systematic errors originating from the vertex reconstruction are
estimated by modifying the vertex quality and track quality selection
for the tagging side by $\pm10\%$ and by varying the
$B$ flight length assumed in the
interaction point constraint by $\pm 10~\mu$m.
We also check the result with different $\Delta t$ ranges, $\pm$40 ps or
$\pm$100 ps instead of the nominal range of $\pm$70 ps.
The resolution function uncertainty is obtained by modifying each
parameter in the resolution function by $\pm 1 \sigma$.
The dependence on the $B^0$ lifetime and the $\Delta m_d$ is measured by
varying the measured values by $\pm 1 \sigma$.
We test for a bias in a reconstruction
with large statistics signal MC samples and
observe no statistically significant discrepancy, therefore no systematic error
due to reconstruction bias is included.

The total effective efficiency obtained by summing over the six $r$ regions is
\[
\epsilon_{\rm eff} =
\sum_l \epsilon_l(1-2w_l)^2 = (28.8 \pm 0.6)\%,
\]
where $\epsilon_l$ is the event fraction in each of the six regions. The
error includes both statistical and systematic contributions.

\section{Summary}

We have developed a flavor tagging algorithm for the measurement of
the $CP$ violation parameter $\sin 2 \phi_1$ and other measurements
at the Belle experiment.
The algorithm is designed to maximize the effective tagging efficiency,
$\epsilon (1-2w)^2$, where $\epsilon$ is the efficiency and $w$ is
the wrong tag fraction.
We introduce two variables $q$ and $r$, where $q$ is the flavor
charge of a $B$ meson and $r$ is the MC-determined
event-by-event flavor dilution factor.
The value $r$ is related to $w$ through $r = 1-2 w$.
In our approach, we independently determine $w$ from control data samples,
which are self-tagging events and check the relation, $r = 1-2 w$.
We therefore avoid possible biases that may be introduced by the use of MC.
We have achieved an effective efficiency of $29.3\pm 0.1\%$ according to MC
simulation, which is significant improvement over the value of $22.2\pm 0.1\%$
achieved by the classical method of flavor tagging
using only leptons and kaons.

The flavor tagging performance is estimated from
samples of $B^0$ decays into the self-tagged modes,
$B^0 \rightarrow D^{*-} \ell^+ \nu, D^{*-} \pi^+, D^{*-} \rho^+$ and
$D^- \pi^+$. A total of 65332 events are used to evaluate the performance.
We obtain an effective tagging efficiency of
($28.8 \pm 0.6$)\%, which agrees well with the MC expectation.

\section*{Acknowledgments}
We would like to thank the members of the Belle collaboration.
We especially acknowledge those who calibrate the tracking
and the particle identification devices of the Belle detector.
We are grateful to T.~E.~Browder for a careful reading of
the manuscript.
This work was supported in part by Grant-in-Aid for Scientific Research
on Priority Areas (Physics of $CP$ violation) from
Ministry of Education, Culture, Sports, Science and Technology of Japan.


\begin{table}[p]
\begin{center}
\caption{Discriminants for the electron-like and muon-like track categories}
\label{table:lepton}
\begin{tabular}{cm{85mm}c} \hline \hline
variable & description & number of bins \\ \hline
charge & track charge & 2 \\
$e$ or $\mu$ & identifier of an electron or muon & 2 \\
lepton ID & lepton-ID quality value & 4 \\
$p^{\rm cms}_\ell$ & the magnitude of the momentum in the CMS & 11 \\
$\theta_{\rm lab}$ & the polar angle in the laboratory frame & 6 \\
$M_{\rm recoil}$ & the hadronic recoil mass & 10 \\
$P^{\rm cms}_{\rm miss}$ & the magnitude of the missing momentum in
the CMS & 6 \\
\hline
total & & 31680 \\
\hline \hline
\end{tabular}
\end{center}
\end{table}

\begin{table}[p]
\begin{center}
\caption{Discriminants for the slow-pion-like track category}
\label{table:slowpi}
\begin{tabular}{cm{85mm}c} \hline \hline
variable & description & number of bins \\
\hline
charge & track charge & 2 \\
$p^{\rm lab}$ & the magnitude of the momentum in the laboratory frame
& 10 \\
$\theta_{\rm lab}$ & the polar angle in the laboratory frame & 10 \\
$\cos\alpha_{\rm thr}$ & the cosine of the angle between the slow pion
candidate and the thrust axis of the tag-side particles in the CMS & 7 \\
$\pi /e$ ID & pion ID probability from \dedx & 5 \\ \hline
total & & 7000 \\ \hline \hline
\end{tabular}
\end{center}
\end{table}

\begin{table}[p]
\begin{center}
\caption{Discriminants for the kaon-like track category}
\label{table:kaon}
\begin{tabular}{cm{75mm}c} \hline \hline
variable & description & number of bins \\
\hline
charge & Track charge & 2 \\
w/ or w/o $K_S^0$ &
switch to indicate whether the event contain $K_S^0$'s or not & 2 \\
$p^{\rm cms}$ & momentum in the center-of-mass system of $\Upsilon(4S)$ & 21 \\
$\theta^{\rm lab}$ & polar angle in the laboratory system & 18 \\
$K/\pi$ ID & quality value of $K/\pi$ ID(\dedx, TOF, ACC)& 13 \\ \hline
total & & 19656 \\ \hline \hline
\end{tabular}
\end{center}
\end{table}

\begin{table}[p]
\caption{Discriminants for the $\Lambda$-like particle category}
\label{table:lambda}
\begin{center}
\begin{tabular}{cm{75mm}c} \hline \hline
variable & description & number of bins \\
\hline
flavor & flavor of $\Lambda$ ($\Lambda$ or $\bar{\Lambda}$) & 2 \\
$K_S^0$ presence
& switch that indicates whether the event contains $K_S^0$'s or not & 2 \\
$M_{p \pi}$ &
invariant mass of the pion and the proton candidate at the secondary vertex
& 2 \\
$\theta_{defl}$ &
the angle difference between the $\Lambda$ momentum vector and the direction
of the $\Lambda$ vertex point from the nominal IP & 2 \\
$\Delta z$ & $z$ difference of the two tracks at the $\Lambda$ vertex point
& 2 \\ \hline
total & & 32 \\ \hline \hline
\end{tabular}
\end{center}
\end{table}

\begin{table}[p]
\begin{center}
\caption{Discriminants for event-level tagging}
\label{table:event}
\begin{tabular}{cm{85mm}c} \hline \hline
variable & description & number of bins \\ \hline
$(q \cdot r)_{l}$ & $(q \cdot r)$ for the highest $r$ from outputs of
lepton category & 25 \\
$(q \cdot r)_{K/\Lambda}$ & $q \cdot r$ from the product of likelihood;
$\frac{\prod_i[1+(q \cdot r)_i] - \prod_i[1-(q \cdot r)_i]}
{\prod_i[1+(q \cdot r)_i] + \prod_i[1-(q \cdot r)_i]}$
where the subscript $i$ runs over all outputs of the kaon and $\Lambda$
categories. & 35 \\
$(q \cdot r)_{\pi_s}$ & $q \cdot r$ for the highest $r$ from  outputs of
slow pion category & 19 \\ \hline
total & & 16625 \\ \hline \hline
\end{tabular}
\end{center}
\end{table}

\begin{table}[p]
\begin{center}
\caption{Event fractions $\epsilon_l$,
\label{tag_sum}
wrong tag fractions $w_l$, and effective tagging efficiencies
$\epsilon_{\rm eff}^l = \epsilon_l(1-2w_l)^2$ for each $r$ interval.
The first errors and second errors of $w_l$ are statistical and systematic
uncertainties, respectively. The errors of $\epsilon_{\rm eff}^l$ are
statistical and systematic combined.
The event fractions are obtained from the {$\jpsi\ks$} simulation.}
\begin{tabular}{cccll} \hline \hline
$l$ & $r$ interval & $\epsilon_l$
&\multicolumn{1}{c}{$w_l$} & \multicolumn{1}{c}{$\epsilon_{\rm eff}^l$} \\ \hline
1 & 0.000 -- 0.250 & 0.398 & $0.458\pm0.005\pm0.003$ & $0.003\pm0.001$ \\
2 & 0.250 -- 0.500 & 0.146 & $0.336\pm0.008\pm0.004$ & $0.016\pm0.002$ \\
3 & 0.500 -- 0.625 & 0.104 & $0.228\pm0.009~^{+0.004}_{-0.006}$ & $0.031\pm0.002$ \\
4 & 0.625 -- 0.750 & 0.122 & $0.160\pm0.007~^{+0.005}_{-0.004}$ & $0.056\pm0.003$ \\
5 & 0.750 -- 0.875 & 0.094 & $0.112\pm0.008\pm0.004$ & $0.056\pm0.003$ \\
6 & 0.875 -- 1.000 & 0.136 & $0.020~^{+0.005}_{-0.004}~^{+0.005}_{-0.004}$ & $0.126~^{+0.003}_{-0.004}$ \\ \hline \hline
\end{tabular}
\end{center}
\end{table}

\begin{table}[p]
\begin{center}
\caption{Wrong tag fractions
$w_l$ for $B^0$ tagged ($q=+1$) and $\bzb$ tagged ($q=-1$) events separately.
The error is statistical only.}
\label{q_pm_diff}
\begin{tabular}{ccll} \hline \hline
$l$ & $r$ interval & $w_l$ for $q=+1$ & $w_l$ for $q=-1$ \\ \hline
1 & 0.000 -- 0.250 & $0.462\pm0.007$ & $0.453\pm0.007$ \\
2 & 0.250 -- 0.500 & $0.339\pm0.011$ & $0.333\pm0.011$ \\
3 & 0.500 -- 0.625 & $0.211\pm0.012$ & $0.246~^{+0.013}_{-0.012}$ \\
4 & 0.625 -- 0.750 & $0.148\pm0.010$ & $0.173\pm0.011$ \\
5 & 0.750 -- 0.875 & $0.101\pm0.011$ & $0.122\pm0.011$ \\
6 & 0.875 -- 1.000 & $0.020~^{+0.007}_{-0.006}$ & $0.020\pm0.006$ \\ \hline \hline
\end{tabular}
\end{center}
\end{table}

\begin{table}[p]
\begin{center}
\caption{Summary of systematic errors ($\times 10^{-2}$).}
\label{sys_sum}
\begin{tabular}{lllllll} \hline \hline
source & $~~~w_1$ & $~~~w_2$ & $~~~w_3$ & $~~~w_4$ & $~~~w_5$
& $~~~w_6$ \\ \hline
Semileptonic signal fraction & $\pm0.17$ & $\pm0.26$ & $\pm0.30$
& $~^{+0.26}_{-0.25}$ &$~^{+0.31}_{-0.29}$ & $~^{+0.19}_{-0.18}$ \\
Semileptonic background shape & $\pm0.07$ & $~^{+0.11}_{-0.12}$ &
$\pm0.14$ & $\pm0.12$ & $\pm0.14$ & $~^{+0.32}_{-0.10}$ \\
Semileptonic $D^{**}$ composition & $~^{+0.22}_{-0.13}$ &
$~^{+0.13}_{-0.01}$ & $~^{+0.09}_{-0.10}$
& $~^{+0.18}_{-0.10}$ & $~^{+0.14}_{-0.11}$ & $~^{+0.12}_{-0.06}$ \\
Semileptonic $B^+$ background $w$ & $~^{+0.12}_{-0.13}$ & $\pm0.09$ &
$\pm0.09$ & $\pm0.07$ & $~^{+0.07}_{-0.08}$ & $\pm0.03$ \\
Hadronic signal fraction & $~^{+0.03}_{-0.02}$ & $~^{+0.04}_{-0.10}$ &
$~^{+0.08}_{-0.04}$ & $~^{+0.11}_{-0.03}$ & $~^{+0.11}_{-0.04}$ & $\pm0.02$ \\
Hadronic background shape  & $\pm0.01$ & $\pm0.01$  & $\pm0.02$
 & $\pm0.01$ & $\pm0.01$ & $<0.01$ \\
Hadronic background mixing & $~^{+0.04}_{-0}$ & $<0.01$ & $~^{+0.05}_{-0}$
& $~^{+0.03}_{-0}$ & $~^{+0.08}_{-0}$ & $~^{+0.04}_{-0}$ \\
Vertex reconstruction & $~^{+0.08}_{-0.10}$ & $~^{+0.09}_{-0.27}$ &
$~^{+0.13}_{-0.44}$
& $~^{+0.27}_{-0.21}$ & $~^{+0.05}_{-0.13}$ & $~^{+0.09}_{-0.22}$ \\
Resolution parameters & $~^{+0.02}_{-0.01}$ & $~^{+0.03}_{-0.02}$ &
$\pm0.04$
& $~^{+0.04}_{-0.03}$ & $~^{+0.05}_{-0.04}$ & $\pm 0.06$ \\
$B^0$ lifetime and $\Delta m_d$ & $\pm0.02$ & $~^{+0.10}_{-0.09}$ &
$\pm0.15$
& $~^{+0.17}_{-0.16}$ & $~^{+0.20}_{-0.19}$ & $~^{+0.22}_{-0.20}$ \\ \hline
Total & $\pm 0.3$ & $\pm 0.4$ & $~^{+0.4}_{-0.6}$
& $~^{+0.5}_{-0.4}$ & $\pm 0.4$ &  $~^{+0.5}_{-0.4}$ \\ \hline \hline
\end{tabular}
\end{center}
\end{table}


\appendix

\section{Control Samples for Flavor Tagging}
\label{sec:ctrl_samples}
Control samples of self-tagged $B$-mesons are used to measure $w$'s
directly from data, and minimize the systematic uncertainty in our
$\sin 2 \phi_1$ measurement. The control samples are also used for the
evaluation of the flavor tagging performance $\epsilon_{\rm eff}$
and to check flavor tagging inputs and outputs.
We use the semileptonic decay mode,
$B^0 \rightarrow D^{*-}\ell^+\nu$\cite{dslnu_mixing}
and hadronic modes $B^0 \rightarrow D^{(*)-}\pi^+$,
and $D^{*-}\rho^+$\cite{hadron_mixing}
and their charge conjugates as control samples.
We fully reconstruct those $B$-meson decays
and tag the $b$-flavor of the associated $B$-mesons
using the algorithm described in Section~\ref{sec:algorithm}.
Using the 78~fb$^{-1}$ data sample corresponding to
$85 \times 10^{6}$ $B\overline{B}$ pairs,
we select 47317 semileptonic decay candidates with 79.2\% purity and 18015
hadronic decay candidates with purities of 87.9\%, 84.3\% and 72.6\%
for the $D^-\pi^+$, $D^{*-}\pi^+$ and $D^{*-}\rho^+$ modes, respectively.
In the $w$ measurement and $\epsilon_{\rm eff}$ evaluation described in
Section~\ref{sec:performance}, the effect of $B^0$-$\bzb$ mixing
is taken into account by fitting the time dependence of the flavor mixing.
For the figures that show flavor tagging input variables
and outputs $q\cdot r$ in Section~\ref{sec:algorithm},
we use $\chi_d = 0.181 \pm 0.004$\cite{PDG2002} to take the effect of mixing
into account. In those figures, the number of entries in an event
for $\bzb$ ($B^0$) is calculated as
$\frac{N_{OF} \cdot \{1-\chi_d\} - N_{SF} \cdot \chi_d}
{n_{OF} \cdot \{1-\chi_d\} - n_{SF} \cdot \chi_d}$,
where $N_{OF}$ and $n_{OF}$ are the entries and the total number of events
from the background subtracted $D^{(*)-} X^+$ ($D^{(*)+} X^-$) control samples,
respectively,
and $N_{SF}$ and $n_{SF}$ are those from the
background subtracted $D^{(*)+} X^-$ ($D^{(*)-} X^+$) control samples.



\end{document}